\documentclass[a4paper,11pt]{article}
\pdfoutput=1 % if your are submitting a pdflatex (i.e. if you have images in pdf, png or jpg format)

\usepackage{jheppub} 

\usepackage[T1]{fontenc} % if needed

\usepackage{amsmath,amssymb}
\usepackage{mathtools}
\usepackage[usenames, dvipsnames]{xcolor}
\usepackage{braket}
\usepackage{graphicx,bm}
\usepackage{slashed}
\usepackage{tensor,siunitx,bbold}

\usepackage[normalem]{ulem}  % \sout{old text} for strikeout

\renewcommand\sout{\bgroup \color{red} \ULdepth=-.5ex \ULset}

\title{\boldmath
Analysis of the $b_1$ meson decay in local tensor bilinear representation}

\author[a]{Kie Sang Jeong}

\author[b]{Su Houng Lee}

\author[c,a]{Yongseok Oh}

\affiliation[a]{Asia Pacific Center for Theoretical Physics, Pohang, Gyeongbuk 37673, Korea }

\affiliation[b]{Department of Physics and Institute of Physics and  Applied Physics, Yonsei University, Seoul 03722, Korea}

\affiliation[c]{Department of Physics, Kyungpook National University, Daegu 41566, Korea}

\emailAdd{kiesang.jeong@apctp.org}
\emailAdd{suhoung@yonsei.ac.kr} 
\emailAdd{yohphy@knu.ac.kr}  

\abstract{
We explore the validity of vector meson dominance in the radiative decay of the $b_1(1235)$ meson.
In order to explain the violation of the vector meson dominance hypothesis in this decay process, 
we investigate a model where the $b_1$ meson strongly couples with the local current in tensor bilinear representation.
The tensor representation is investigated in the framework of the operator product expansion and we found
a low energy decay process that does not follow the usual vector meson dominance hypothesis.
The $\omega$-like intermediate meson state of quantum numbers $I^G(J^{PC}) = 0^{-}(1^{--})$ is found to have a nontrivial role in
the decay process of the $b_1$ meson. 
The spectral structure of the $\omega$-like state is found to be close to a $\pi$-$\rho$ hybrid state, which provides
a mechanism that evades the usual vector meson dominance hypothesis. 
Precise measurements of various decay channels of the $b_1$ meson are, therefore, required to unravel the internal structure of
axial vector mesons.
}

\begin{document} 
\maketitle
\flushbottom

\section{Introduction}
\label{sec:intro}

According to the Particle Data Group~\cite{PDG16}, the axial-vector $b_1(1235)$ meson has a mass 
of $1229.5 \pm 3.2~\textrm{MeV}$ and quantum numbers $I^G(J^{PC}) = 1^+ (1^{+-})$.
Its decay width is estimated to be $\Gamma(b_1) = 142 \pm 9~\textrm{MeV}$, which is dominated by
the $b_1 \to \pi \omega$ decay channel.
In particular, the $D$-wave to $S$-wave amplitude ratio in the decay of $b_1 \to \pi \omega$ is found to be 
$D/S = 0.277 \pm 0.027$.%
\footnote{The $b_1 \to \pi\rho$ decay violates $G$-parity conservation.
The most recent measurement performed at the Brookhaven National Laboratory gives 
$D/S = 0.269 \pm (0.009)_{\rm stat} \pm (0.01)_{\rm sys}$~\cite{E852-02}.}
The radiative decays of the $b_1(1235)$ meson were also observed and the width is estimated as
$\Gamma(b_1\to \pi^{\pm} \gamma) = 230 \pm 60~\textrm{keV}$~\cite{CHJM84, PDG16}.%
\footnote{The radiative decay of $b_1 \to \pi^0 \gamma$ is yet to be measured.}
Since most strong decays of the $b_1$ meson is in the $\pi\omega$ channel, its radiative decay into $\pi \gamma$
can be another test for the vector meson dominance (VMD) hypothesis.

If the U(1) symmetry part of the vector meson field is conserved, the vector meson field plays 
the role of the external electromagnetic field source through the direct
coupling in the form of $\mathcal{L}_{\gamma\omega} \sim A_\mu V^\mu$~\cite{LWZ67,KLZ67,GS68, SS72a,Feynman-PHI}. 
This argument is based on the assumption that the vector meson has the same representation
as the electromagnetic current.%
\footnote{The realization of VMD in effective Lagrangian approaches can be found, for example, in
Refs.~\cite{ZB86, Meissner88,FKTU85,BKY88}.}
According to the VMD hypothesis, it is natural to assume that the radiative $b_1$ decay is dominated by the 
intermediate $\omega$ meson, namely, it leads to the decay chain of
$b_1\rightarrow\pi^{\pm}\omega \rightarrow \pi^{\pm}\gamma$.
However, as will be discussed in Sec.~\ref{sec:phenom}, such VMD model underestimates the decay width of
$b_1 \to \pi^{\pm}\gamma$~\cite{Rosner81b,IYO89}, and this may imply the possibility
that there might be a missing source or mechanism for the radiative decays of the $b_1$ meson, 
which dominates over the mechanism proposed by the VMD hypothesis. 
The present work is motivated by the idea that, if the intermediate spin-1 mesonic state is not in exactly the same representation 
with the electromagnetic current, then there could be an additional process for the $b_1$ radiative decay
that cannot be ascribed to the VMD hypothesis alone. 
In other words, another representation for spin-1 mesonic states can be allowed 
if it overlaps with the U(1) gauge boson in low energy regime where
flavor symmetry is broken down to $\mbox{SU(2)}_{V}$~\cite{CZ84, CJ96}.
In this case, the state with the quantum numbers of the $\omega$ meson, usually represented by chiral vector current, 
can also be represented in the helicity mixed space via a tensor current. 
We denote this state as $\bar{\omega}$.
In this helicity mixed representation, the $b_1\rightarrow\pi \bar\omega$ decay can be regarded as a production process of a soft-pion 
as the $b_1$ and $\bar\omega$ states are chiral partners to each other.

In this study, the properties of the tensor representation for spin-1 mesonic current is reviewed 
and their current correlation function is calculated in the framework of operator product expansion
(OPE)~\cite{WZ72}. 
We first discuss the spin-1 mesonic state interpolated by tensor currents and the $b_1$ decay 
process in the context of chiral rotation~\cite{Gell-Mann62,GOR68}. 
The distinct nature of the spin-1 mesonic state in the tensor current correlation is analyzed through 
QCD sum rules~\cite{SVZ79a,SVZ79b,SVZ79c,RRY85,GRDW87,BB96}. 
In the OPE of tensor current correlator, the four-quark condensates appear as the leading 
quark contribution. 
Although the vacuum saturation hypothesis for the four-quark condensate has been widely used in
previous studies, it does not reflect all the possible vacuum structure of QCD. 
The factorization scheme is modified to take into account the non-factorizable part and quark 
correlation pattern in topologically nontrivial gauge 
configuration~\cite{CD81,BC80,LS92b,EHS96a,Cohen96,LH96,LC13}. 
Within the usual limit of the factorization parameters, we will show that the tensor current can strongly couple 
to the $b_1$ meson as well as to the $\pi$-$\rho$ continuum state that has the quantum numbers of the $\omega$ meson.
 Analyzing the $b_1 \to \pi \bar\omega$ decay within the soft pion limit, we find that the $\omega$-like
mesonic state $\bar\omega$, which appears in the strong decay of the $b_1$ into the pion, would not be a simple one-particle state 
but an anomalous hybrid state.
Our analyses will show that this process would be a source for $b_1$ radiative decays instead of the usual VMD
mechanism.

This paper is organized as follows.
In section~\ref{sec:phenom}, we discuss the problems in the phenomenology of the $b_1$ meson decays.
In section~\ref{sec:tensor}, a brief review on local bilinear representations of spin-1 mesonic state and
their evolution in chiral symmetry breaking is presented. 
Detailed OPE for correlation functions and corresponding spectral analyses with 
four-quark condensates are presented in section~\ref{sec:ope}.
Section~\ref{sec:summary} contains discussion and conclusions.
We leave the details on tensor currents and technical remarks on sum rule analysis in Appendixes.

\section{\boldmath Phenomenology of the $b_1$ meson}
\label{sec:phenom}

In this Section, we review the phenomenology of the $b_1$ meson decays.
The major strong decay channel, i.e., $b_1 \to \pi\omega$, is discussed and then its
radiative decays is discussed in connection with the VMD hypothesis.

\subsection{Strong decay}

In quark models, the $b_1(1235)$ state is described as a member of the ${}^1P_1$ nonet
(in the ${}^{2S+1}L_J$ notation), where the $q\bar{q}$ pair is in the relative $P$-wave state,
having quantum numbers $J^{PC} = 1^{+-}$.
Other members of this nonet may include the $h_1(1170)$ and $h_1(1380)$ which are iso-scalars, 
and $K_1(1270)$ or $K_1(1400)$ that has one unit of strangeness~\cite{IYO89}.

The strong decay of the $b_1$ meson is described by the amplitude of an axial-vector meson 
($J^P = 1^+$) decay into a vector meson ($J^P = 1^-$) plus a pseudoscalar meson ($J^P = 0^-$),
which reads
\begin{equation}
\mathcal{M}(1^+ \to 1^- 0^-) = \varepsilon_\nu^*(1^-,p_V^{})
\Gamma^{\mu\nu}(1^+ \to 1^-0^-) \varepsilon_\mu (1^+, p_A^{}),
\end{equation}
where $\varepsilon_\mu (1^\pm, p)$ represents the polarization vector of $J^P = 1^\pm$ meson
with momentum $p$.
Here, $p_A^{}$ and $p_V^{}$ are the momenta of the axial-vector meson and the vector meson,  
respectively, and the momentum of the pseudoscalar meson $p_P^{}$ is determined by the
energy-momentum conservation, $p_A^{} = p_V^{} + p_P^{}$.
The most general form of the amplitude $\Gamma^{\mu\nu}$ reads~\cite{IMR89}
\begin{eqnarray}
\Gamma^{\mu\nu}(1^+ \to 1^-0^-) &=& f_1^{} g^{\mu\nu} + f_2 (p_V^{} - p_P^{})^\mu 
(p_A^{} + p_P^{})^\nu + f_3 (p_V^{} + p_P^{})^\mu (p_A^{} + p_P^{})^\nu
\nonumber \\ && \mbox{}
+ f_4 (p_V^{} - p_P^{})^\mu (p_A^{} - p_P^{})^\nu + f_5
(p_V^{} + p_P^{})^\mu (p_A^{} - p_P^{})^\nu,
\end{eqnarray}
where $f_i$'s are form factors.
If all of the particles in the decay process are on-mass shell, only the two
terms, $f_1$ and $f_2$, are non-vanishing because of
$\varepsilon(1^-,p_V^{})\cdot p_V^{} = \varepsilon(1^+,p_A^{})\cdot
p_A^{} = 0$.

Therefore, the decay amplitude for $b_1 \to \pi \omega$ constrained by gauge invariance reads~\cite{IMR89}
\begin{equation}
\left\langle \omega(k) \pi | H_{\rm int} | b_1 (q)  \right\rangle
= -i \varepsilon_\nu^* (\omega, k) \Gamma^{\mu\nu}
\varepsilon_\mu(b_1,q),
\end{equation}
where the momenta of the $b_1$ and $\omega$ mesons are represented by $q$ and $k$, respectively, and
\begin{eqnarray}
\Gamma^{\mu\nu} &=& F_{b_1\omega\pi} \left( g^{\mu\nu} - \frac{k^\mu k^\nu}{k^2} -
\frac{q^\mu q^\nu}{q^2} + \frac{k \cdot q}{k^2 q^2} q^\mu k^\nu \right)
\nonumber \\ && \mbox{}
+ G_{b_1\omega\pi} \left( k^\mu - \frac{k \cdot q}{q^2} q^\mu \right)
 \left( q^\nu - \frac{k \cdot q}{k^2} k^\nu \right).
\end{eqnarray}
Because of the transversality condition, $\varepsilon(p) \cdot p=0$, however, the terms with $q^\mu$ or $k^\nu$ vanish,
and we can rewrite it effectively as
\begin{align}
\Gamma^{\mu\nu} = F_{b_1\omega\pi} g^{\mu\nu} + G_{b_1\omega\pi} k^\mu q^\nu.
\label{b1decay1}
\end{align}

By introducing dimensionless form factors, $f$ and $h$, we can write~\cite{OL02}
\begin{equation}
\Gamma^{\mu\nu} = M_{b_1} \left[ f g^{\mu\nu} +
\frac{h}{M_\omega M_{b_1}} q^\nu k^\mu \right],
\label{b1-om-pi}
\end{equation}
where $M_{b_1}$ and $M_\omega$ are the masses of the $b_1$ meson and $\omega$ meson, respectively.
Then the decay width is calculated as
\begin{equation}
\Gamma(b_1 \to \omega\pi) = \frac{|\bm{k}|}{24\pi} \left\{ 2f^2 +
\frac{1}{M_\omega^4} \left( E_\omega M_\omega f + |\bm{k}|^2 h \right)^2
\right\},
\label{b1-decay}
\end{equation}
where $E_\omega$ is the energy of the $\omega$ meson,
$E_\omega = \sqrt{M_\omega^2 + |\bm{k}|^2 }$.
Also from the definitions of the $S$- and $D$-wave amplitudes,
\begin{align}
\left\langle \omega(\bm{k},m_\omega) \pi(-\bm{k}) | H_{\rm int} | b_1 (\bm{0}, m_b) \right\rangle
=&~ i f^S \delta_{m_\omega m_b} Y_{00} (\Omega_k) 
\nonumber \\ & \mbox{}
+ i f^D \sum_{m_\ell}
\left\langle 2 m_\ell 1 m_\omega | 1 m_b \right\rangle Y_{2m_\ell} (\Omega_k),
\end{align}
we have
\begin{align}
f^S =& \frac{\sqrt{4\pi} M_{b_1}}{3M_\omega^2} \left[ M_\omega (
E_\omega + 2M_\omega) f + |\bm{k}|^2 h \right],
\nonumber \\
f^D =& -\frac{\sqrt{8\pi} M_{b_1}}{3M_\omega^2} \left[ M_\omega (
E_\omega - M_\omega) f + |\bm{k}|^2 h \right],
\end{align}
which gives
\begin{equation}
R = f^D/f^S = - \frac{\sqrt2 \left\{ M_\omega ( E_\omega - M_\omega) f + |\bm{k}|^2 h \right\}}
{\left\{ M_\omega (E_\omega + 2M_\omega) f + |\bm{k}|^2 h \right\}}.
\end{equation}
The magnitude of the momentum $\bm{k}$ is 
\begin{equation}
|\bm{k}| = \frac{1}{2M_{b_1}} \sqrt{\lambda(M_{b_1}^2,
M_\omega^2, M_\pi^2)}, 
\end{equation}
where $\lambda(x,y,z) = x^2 + y^2 + z^2 -2xy - 2yz - 2zx$ is the K{\"a}ll{\'e}n function.
This gives $|\bm{k}| = 346.5$~\si{MeV} with $M_{b_1} = 1.230$~GeV, $M_\omega = 0.783$~MeV, 
and $M_\pi = 0.140$~GeV.
From the two experimental data,
\begin{equation}
\Gamma(b_1 \to \omega\pi) \simeq \Gamma(b_1) = 142 \pm 9 \mbox{ MeV},
\qquad f^D/f^S = 0.277 \pm 0.027,
\end{equation}
we obtain
\begin{equation}
f = 3.70, \qquad h = -11.04,
\end{equation}
up to the overall phase, which leads to%
\footnote{Note that this is very different from the old estimation of
Refs.~\cite{Barmawi66a,Barmawi66b,Barmawi68b}, which obtained $h/f \approx +9.0$
by fitting the high energy data of $\pi N \to \omega N$ with
the exchanges of the $b_1$ and $\rho$ trajectories.}
\begin{equation}
h/f = - 2.98.
\end{equation}

One can relate the above information with the couplings of the effective 
$b_1 \omega\pi$ interaction Lagrangian,
\begin{equation}
\mathcal{L} = g_1^{} M_{b_1} \omega_\mu {\bm{b}_1}^{\mu} \cdot \bm{\pi}
+ \frac{g_2^{}}{M_{b_1}} \, \omega_{\mu\nu}\, \bm{b}_1^{\mu\nu} \cdot \bm{\pi},
\label{lag:b1}
\end{equation}
where $\omega^\mu$ and $b_1^\mu$ are the $\omega$ meson field and the $b_1$
meson field, respectively, and their field strength tensors are
\begin{equation}
\omega_{\mu\nu} = \partial_\mu \omega_\nu - \partial_\nu \omega_\mu,
\qquad
b_1^{\mu\nu} = \partial^\mu b_1^\nu - \partial^\nu b_1^\mu,
\end{equation}
and $\bm{b}_1^{\mu} \cdot \bm{\pi} = b_1^0 \pi^0 + b_1^+ \pi^- + b_1^- \pi^+$.%
\footnote{This flavor structure is consistent with the form of $\mbox{Tr}(V[B,P]_+)$, which should
be compared with the axial-vector meson interaction that comes from the structure of $\mbox{Tr}(V[A,P]_-)$.
Here, $V$, $A$, $B$, $P$ are the vector meson octet, axial-vector meson octet including the $a_1^{}$ meson, 
axial-vector meson octet including the $b_1$ meson, and pseudoscalar meson octet and $[A,B]_\pm = AB \pm BA$.}
The above Lagrangian gives
\begin{equation}
\Gamma^{\mu\nu} (b_1 \to \omega \pi) = (g_1^{} + 2 g_2^{} \, k \cdot q) g^{\mu\nu}
- 2 g_2^{} \, k^\mu q^\nu.
\end{equation}
Comparing with Eq.~(\ref{b1decay1}) we have
\begin{eqnarray}
F_{b\omega\pi} &=& M_{b_1} f = M_{b_1} g_1^{} + \frac{2 k \cdot q}{M_{b_1}} \, g_2^{} , 
\nonumber \\
G_{b\omega\pi} &=& \frac{h}{M_\omega}= -\frac{2}{M_{b_1}} g_2^{}.
\end{eqnarray}
Then we can estimate the coupling constants as
\begin{equation}
g_1^{} = -1.34 \qquad g_2^{} = 8.61.
\end{equation}

\subsection{Radiative decay and vector meson dominance}

The radiative decay of $b_1 \to \pi\gamma$ was investigated in quark models based on the VMD hypothesis.
The VMD hypothesis leads to
\begin{equation}
\Gamma(b_1 \to \gamma\pi) = \left(\frac{|\bm{p}_\gamma^{}|}{|\bm{p}_\pi^{}|}\right)^n
\frac{\alpha_{\rm em}}{g_\omega^2/4\pi} \Gamma(b_1 \to \omega_\perp^{} \pi),
\label{vmd1}
\end{equation}
where $\omega_\perp^{}$ represents the transverse component of the $\omega$ vector meson
and $g_\omega$ is the coupling of the $\omega$ meson to the photon in VMD that is estimated as
$g_\omega \approx 17.14$. 
The magnitudes of the momenta of the decays are $|\bm{p}_\gamma^{}| = 607.0$~MeV and 
$|\bm{p}_\pi^{}| = 346.5$~MeV.
Here, $n$ shows the characteristic momentum dependence of the decay process, of which
values will be discussed later.

The decay amplitude of the strong decay of the $b_1$ into the transverse $\omega$ meson can be 
estimated following Ref.~\cite{Rosner81b}.
The transverse and longitudinal decay amplitudes of the $b_1 \to \omega \pi$ decay are related to
the $S$ and $D$ wave amplitudes as
\begin{equation}
A_1 = f^S + \frac{1}{\sqrt2} f^D, \qquad A_0 = f^S - \sqrt2 f^D,
\end{equation}
which gives
\begin{equation}
A_1 = \sqrt{4\pi} M_{b_1} f, \qquad
A_0 = \frac{\sqrt{4\pi}M_{b_1}}{M_\omega^2} \left( M_\omega E_\omega f +
|\bm{k}|^2 h \right).
\end{equation}
So we have
\begin{equation}
\frac{\Gamma(b_1\to\omega_\perp^{}\pi)}{\Gamma(b_1\to\omega\pi)} = \frac{2
|A_1|^2}{2|A_1|^2 + |A_0|^2} = \frac{2 + 2\sqrt{2} R + R^2}{3(1+R^2)}
\simeq 0.8855,
\end{equation}
where the experimental value $R \equiv f^D/f^S = 0.277$ is used. 
Then we have $\Gamma(b_1\to\omega_\perp^{} \pi) \approx 126$~MeV.

In Ref.~\cite{Rosner81b}, $n=3$ is used in the expression of Eq.~(\ref{vmd1}), which gives
\begin{equation}
\Gamma(b_1 \to \gamma \pi) \approx 210~\mbox{keV}.
\end{equation}
This seems to be successful to explain the observed value $\Gamma(b_1 \to \gamma \pi)_{\rm expt.} = 230 \pm 60$~keV.
However, it was pointed out in Ref.~\cite{IYO89} that the VMD prediction is sensitive to the interaction form which determines the
value of $n$.
Using the covariant oscillator quark model, the authors of Ref.~\cite{IYO89} predicted
\begin{equation}
\Gamma(b_1 \to \gamma \pi) \approx 69~\mbox{keV}.
\end{equation}
This is very similar to the value obtained with $n=1$ in Eq.~(\ref{vmd1}).
However, the general form of the $b_1\omega\pi$ interaction contains two terms as shown in Eq.~(\ref{b1-decay}).
In fact, the expression for the decay width (\ref{b1-decay}) contains terms with $n=1$, $3$, and $5$.
Following Ref.~\cite{Rosner81b}, we estimate the radiative decay width of $b_1$ by replacing $\bm{k}$ by $\bm{p}_\gamma$ and
\begin{equation}
f \to \frac{\sqrt{\alpha_{\rm em}}}{g_\omega/\sqrt{4\pi}} f, \qquad
h \to \frac{\sqrt{\alpha_{\rm em}}}{g_\omega/\sqrt{4\pi}} h
\end{equation}
in the decay width formula of Eq.~(\ref{b1-decay}).
Taking into account that $\Gamma(b_1\to\omega_\perp^{} \pi) \approx 126$~MeV, we obtain
\begin{equation}
\Gamma(b_1 \to \gamma \pi) \approx 160~\mbox{keV},
\end{equation}
which is about 2/3 of the measured quantity.

\section{Spin-1 mesons in local tensor bilinear representation}
\label{sec:tensor}

Throughout this study, the isospin flavor matrices are denoted as $T^0= {I}/{2}$, $T^a=\ {\sigma^a }/{2}$, 
where $I$ is the $2 \times 2$ unit matrix and $\sigma^a$ is the Pauli matrix. 
These matrices are normalized as $\mbox{Tr}\, [T^AT^B ]={\delta^{AB}}/{2}$.
Here, capital romans ($A=0-3$) denote isospin ($0\oplus1$) indices and lower-case romans ($a=1-3$) denote isovector indices.
We will use barred romans ($\bar{a}=1-8$) to denote adjoint color indices in gauge interactions.
Mesonic quantum numbers are represented by $\left[ I^{G}(J^{PC})\right]$.

\subsection{Brief review on local bilinear representation of spin-1 mesonic state}

The simplest local representation of the $b_1$ meson is $\bar{q}  T^a \gamma_5^{}
\overleftrightarrow{D}_{\mu}q $. 
This current is in helicity mixed representation $\left(\frac{1}{2} , \frac{1}{2} \right) \oplus \left(\frac{1}{2}, \frac{1}{2}\right)$,
where the first and second numbers in the bracket show the $\mbox{SU(2)}_L$ and $\mbox{SU(2)}_R$ representations,
respectively.  
The derivative leads to additional $p^2$  factors in the Wilson coefficient of OPE terms, which makes the OPE  
dominated by the continuum contribution~\cite{RRY85}. 
In the same representation, tensor current $\bar{q}\, T^a \sigma_{\mu\nu}^{} \, q $, which does not contain the 
covariant derivative, can also be considered. 
This current can couple to four different spin-1 mesonic systems.  
In the non-relativistic limit, where $p_\mu \to (m,\bm{0})$, depending on the intrinsic quantum numbers, the spin-1 mesonic 
states of momentum $p$ and polarization $\lambda$ have the following relations:~\cite{CJ96}
\begin{align}
\bar{\omega}:\ [I^GJ^{PC} = 0^{-}(1^{--})] &
\to \braket{ 0 \mid \bar{q} \, T^0 \sigma_{0k}^{} q \mid \bar{\omega} (p,\lambda) } 
%\nonumber \\ &
= if^T_{\omega} ( -\bar{\epsilon}_k^{(\lambda)} p_0^{} )\label{n1},
\\
\bar{h}_1:\ [I^GJ^{PC} = 0^{-}(1^{+-})] & 
\to \braket{ 0 \mid \bar{q}\,  T^0 \sigma_{ij}^{} q  \mid \bar{h}_1 (p,\lambda) } 
%\nonumber \\ &
= if^T_{h_1}\epsilon_{ijk}^{} ( - \bar{\epsilon}_k^{(\lambda)} p_0^{} )
\nonumber \\ &
\rightarrow \frac{1}{2} \epsilon_{ijk}^{} 
\braket{ 0 \mid \bar{q} \, T^0 \sigma_{ij} q  \mid \bar{h}_1 (p,\lambda) } 
%\nonumber \\ &
= if^T_{h_1}  ( -\bar{\epsilon}_k^{(\lambda)} p_0^{} )
\label{n2},\\
\bar{\rho}:\ [I^GJ^{PC} = 1^{+}(1^{--})] & 
\to \braket{ 0 \mid \bar{q} \, T^a \sigma_{0k}^{} q  \mid \bar{\rho} (p,\lambda) }
%\nonumber \\ &
= if^T_{\rho^a} ( - \bar{\epsilon}_k^{(\lambda)} p_0^{} )
\label{n3},\\
\bar{b}_1:\ [I^GJ^{PC} = 1^{+}(1^{+-})]& \rightarrow 
\braket{ 0 \mid \bar{q}  T^a \sigma_{ij}^{} q  \mid \bar{b}_1 (p,\lambda) } 
%\nonumber \\ &
= if^T_{b_1^a}\epsilon_{ijk}^{}  ( - \bar{\epsilon}_k^{(\lambda)} p_0^{} )
\nonumber\\
& \rightarrow  \frac{1}{2} \epsilon_{ijk}^{}  
\braket{ 0 \mid \bar{q} \, T^a \sigma_{ij}^{} q  \mid \bar{b}_1 (p,\lambda) } 
%\nonumber \\ &
= if^T_{b_1^a}  ( - \bar{\epsilon}_k^{(\lambda)} p_0^{} )
\label{n4},
\end{align}
where the bar notation is used \textit{to distinguish the mesonic states of the helicity mixed tensor current representation
from those of chiral vector representation\/}.%
\footnote{If we extend this formalism to flavor SU(3), there should be states of open and hidden strangeness in tensor representation~\cite{CJ96}.
In the quark model, the $K_1(1270)$ and $K_1(1400)$ are interpreted as mixtures of ${}^1P_1$ and ${}^3P_1$ states~\cite{IYO89}.
Since the $b_1$ is a member of the ${}^1P_1$ multiplet in the quark model, both $K_1$ states may be mixed states of vector and tensor representations.
More rigorous investigation is required to understand the SU(3) realization of tensor representation, but is beyond the scope of this work.}
For example, the $\bar{\omega}$ is the state of tensor current representation having $I^G J^{PC} = 0^- (1^{--})$
as the usual $\omega$ meson.
Here, $\bar{\epsilon}_{\mu }^{(\lambda)}$ is the polarization vector and
the vector meson coupling is given by $f^T_V$ in tensor representation.
In a boosted frame, $p_\mu\rightarrow (\sqrt{m^2 + \bm{p}^2},\bm{p})$, the covariant generalizations read
\begin{align}
\bar{\omega}& \to 
\braket{0 \mid \bar{q} \, T^0 \sigma_{\mu\nu}^{} q  \mid \bar{\omega} (p,\lambda) }
%\nonumber \\ &
= if^T_{\omega} (\bar{\epsilon}_\mu^{(\lambda)} p_\nu^{} -
\bar{\epsilon}_\nu^{(\lambda)} p_\mu^{} ),
\\
\bar{h}_1& 
\to -\frac{1}{2} \epsilon_{\mu \nu \alpha \beta}  
\braket{ 0 \mid \bar{q}\, T^0 \sigma^{ \alpha \beta} q \mid \bar{h}_1 (p,\lambda) }
%\nonumber \\ &
= if^T_{h_1} ( \bar{\epsilon}_\mu^{(\lambda)} p_\nu^{} -
\bar{\epsilon}_\nu^{(\lambda)} p_\mu^{} ),
\\
\bar{\rho}& 
\to \braket{ 0 \mid \bar{q}\, T^a \sigma_{\mu\nu} q  \mid \bar{\rho} (p,\lambda) } 
%\nonumber \\ &
= if^T_{\rho^a} ( \bar{\epsilon}_\mu^{(\lambda)} p_\nu^{} -
\bar{\epsilon}_\nu^{(\lambda)} p_\mu^{} ),
\\
\bar{b}_1& 
\to -\frac{1}{2} \epsilon_{\mu \nu \alpha \beta}^{}
\braket{ 0 \mid \bar{q}\, T^a \sigma^{ \alpha \beta} q \mid \bar{ b}_1 (p,\lambda) } 
%\nonumber \\ &
= if^T_{b_1^a} (\bar{\epsilon}_\mu^{(\lambda)} p_\nu^{} -
\bar{\epsilon}_\nu^{(\lambda)} p_\mu ),
\label{c4}
\end{align}
The tensor current couples only to the transverse polarization of spin-1 mesonic
state~\cite{CZ84, GRDW87, BB96} because of the relation,
\begin{align}
\left\langle 0  \left\vert \bar{q} T^A\sigma_{\mu\nu} q  \right\vert
 [1^{--}] (p,\lambda) \right\rangle &= i f_{V^A}^T \left\{ \bar{\epsilon}_\mu^{(\lambda)}
p_\nu - \bar{\epsilon}_\nu^{(\lambda)} p_\mu \right\}
\nonumber\\
&=i f_{V^A}^T \left\{
\left(g_{\mu\bar{\mu}}-p_{\mu}p_{\bar{\mu}}/p^2\right)
{\bar{\epsilon}^{(\lambda)\bar{\mu}}} p_\nu -
\left(g_{\nu\bar{\nu}}-p_{\nu}p_{\bar{\nu}}/p^2 \right){\bar{\epsilon}^{(\lambda)\bar{\nu}}}  p_\mu \right\}
\nonumber\\
&=i f_{V^A}^T \left\{ \bar{\epsilon}_{\mu }^{t(\lambda)} p_\nu -
\bar{\epsilon}_{\nu }^{t(\lambda)} p_\mu \right\},
\end{align}
where $\bar{\epsilon}_{\mu }^{t(\lambda)}$ denotes the transverse polarization vector.
However, this current couples to both the parity-even and parity-odd modes, and 
each parity mode can be separately projected out from the current-current correlation as
\begin{align}
\sum_\lambda \braket{ 0  \mid \bar{q} T^A \sigma_{\mu\bar{\mu}}^{} q \mid [1^{--}] (p,\lambda) }
\braket{ [1^{--}] (p,\lambda) \mid \bar{q} T^B \sigma_{ \nu \bar{\nu}}^{} q  \mid 0 }
& = -  \delta^{AB} \left({f^T_{-}}\right)^2   p^2  P^{(-)}_{\mu \bar{\mu}, \nu \bar{\nu}},\\
\sum_\lambda \braket{ 0  \mid \bar{q} T^A \sigma_{\mu \bar{\mu}}^{} q \mid [1^{+-}] (p,\lambda) } 
\braket{ [1^{+-}] (p,\lambda) \mid \bar{q} T^B \sigma_{\nu \bar{\nu}}^{} q  \mid 0 }  
& = -  \delta^{AB} \left({f^T_{+}}\right)^2 p^2 P^{(+)}_{\mu \bar{\mu} ,\nu \bar{\nu}},
\end{align}
which defines the projection operators as
\begin{align}
P^{(-)}_{\mu \bar{\mu}; \nu \bar{\nu}} &= g_{\mu\nu}^{} \frac{p_{\bar{\mu}}^{} p_{\bar{\nu}}^{}}{p^2} +
g_{\bar{\mu}\bar{\nu}}^{} \frac{p_{\mu}^{} p_{\nu}^{}}{p^2} - g_{\bar{\mu}\nu}^{} \frac{p_{\mu}^{} p_{\bar{\nu}}^{}}{p^2}
- g_{\mu\bar{\nu}}^{} \frac{p_{\bar{\mu}}^{} p_{\nu}^{}}{p^2},\\
P^{(+)}_{\mu \bar{\mu}; \nu \bar{\nu}} &= P^{(-)}_{\mu \bar{\mu}, \nu \bar{\nu}}
+ \left( g_{\mu\bar{\nu}}^{} g_{\bar{\mu}\nu}^{} - g_{\mu\nu}^{} g_{\bar{\mu}\bar{\nu}}^{} \right),
\end{align}
with the relation $[P^{(\pm)}P^{(\mp)}]_{\mu \bar{\mu}; \nu \bar{\nu}} =0$. 
Considering the assumed symmetries, the four mesonic modes reside in the same  
symmetry group $\mbox{U(2)}_L \times \mbox{U(2)}_R$.  
The group structure allows the transformation among these modes depending on quantum numbers.
For example, both the $\bar\omega$ state and the $\bar{h}_1$ state have isospin $0$ and they can be transformed to each other
through the $\mbox{U}_A(1)$ transformation. 
The same is true for the isospin-1 states, the $\bar\rho$ and the $\bar{b}_1$. 
Similarly, the chiral partner states can be transformed to each other
through the axial $\mbox{SU(2)}$ flavor rotation~\cite{CJ96}, namely, the pair of $\bar{\omega}$ and the $\bar{b}_1$, 
and the pair of the $\bar{\rho}$ and the $\bar{h}_1$.

If all the symmetries are conserved, there should be no correlation between the two different representations of mesonic states. 
However, when chiral symmetry is broken at low energy regime, the mesonic states can have overlaps. 
Since the vector (or chiral) representations of the $\omega$ and $\rho$ states read
\begin{align}
\omega:\ [I^GJ^{PC} = 0^{-}(1^{--})] &
\to \braket{ 0 \mid \bar{q} \, T^0 \gamma_\alpha^{} q  \mid \omega (p,\lambda) } 
= f^V_{\omega} \epsilon_\alpha^{(\lambda)},
\\
\rho:\ [I^GJ^{PC} = 1^{+}(1^{--})] & 
\to \braket{ 0 \mid \bar{q} \, T^a \gamma_\alpha^{} q  \mid \rho (p,\lambda) } 
= f^V_{\rho^a} \epsilon_\alpha^{(\lambda)},
\end{align}
the overlap between the vector and tensor representations can be described as
\begin{align}
& \sum_\lambda 
\braket{ 0  \mid \bar{q}\, T^A \gamma_{\alpha}^{} q \mid [1^{--}] (p,\lambda) } 
\braket{ [1^{--}] (p,\lambda) \mid \bar{q}\, T^B \sigma_{\mu\nu}^{} q \mid 0 } 
\simeq 
-i  \delta^{AB}  f^{V}_{-}  f^T_{-} \mathcal{P}_{\alpha; \mu \nu},
\end{align}
where $P_{\alpha; \mu \nu} = g_{\alpha\mu}^{} p_{\nu}^{} - g_{\alpha\nu}^{} p_\mu^{}$ projects 
only the transverse polarization. 
We also use the polarization sum in the $\mbox{SU(2)}_V$ diagonal phase, i.e.,
$\sum_{\lambda}\epsilon^{(\lambda)}_{\alpha} \bar{\epsilon}^{(\lambda)*}_{\mu} = - g_{\alpha\mu}$. 
Algebraic details for tensor currents and their transformations are presented in Appendix~\ref{appenixb}.

\subsection{$b_1$ meson decay in the soft pion limit}

%%%%%% FIG 1
\begin{figure}
\centering
\includegraphics[height=0.2\columnwidth]{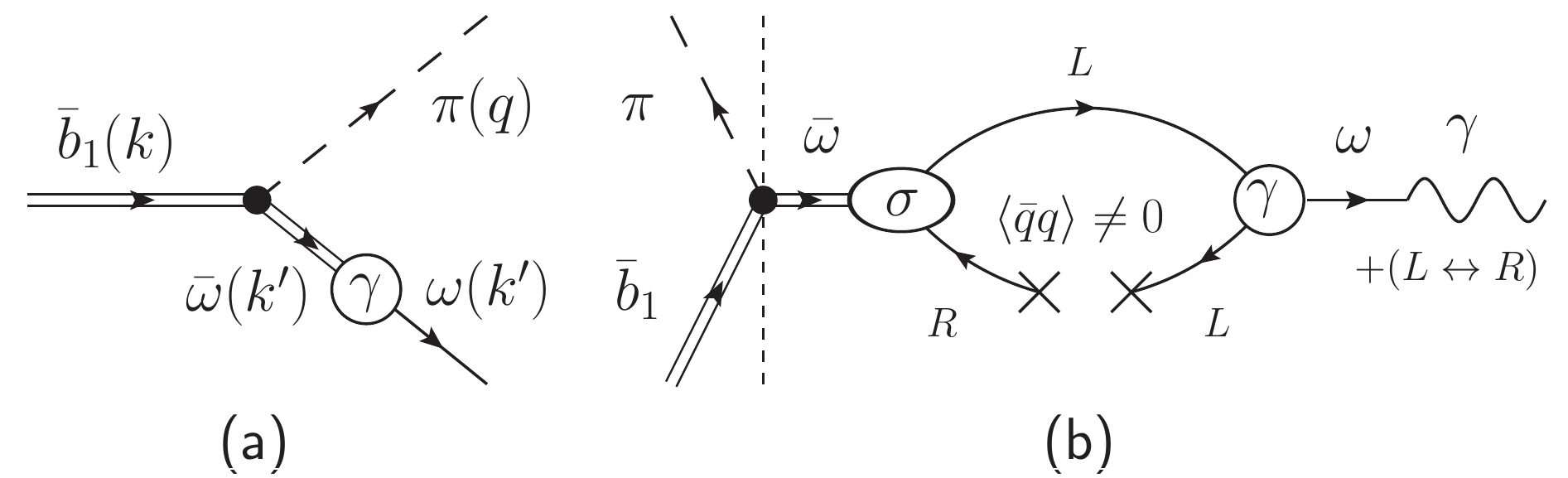}
\caption{The $\bar{b}_1$ decay in chiral representation.
(a) $\bar{b}_1\rightarrow\pi  \bar{\omega}\rightarrow\pi  \omega$ via pion breaking and 
(b) $\bar{b}_1\rightarrow\pi \gamma$ in the VMD hypothesis inferred from Eq.~\eqref{b1gamma}. 
Double lines denote the spin-1 mesonic states in tensor representation. 
The greek letters, `$\sigma$' and `$\gamma$,' in the blob represent the tensor and vector current representation, 
respectively.}
\label{bo}
\end{figure}

The $\bar{b}_1 [1^{+}(1^{+-})]$ and $\bar{\omega}[0^{-}(1^{--})]$ mesonic states transform to each other 
under $ \mbox{SU(2)}_L \times \mbox{SU(2)}_R$ symmetry.
In the $\mbox{SU(2)}_V$ phase, if the $b_1(1235)$ meson state couples strongly to the tensor
current ($\tilde{J}^{a}_{\mu \bar{\mu}}$), the overlap for the decay of $b_1(1235)\rightarrow \pi \omega(782)$ can be expressed as
\begin{eqnarray}
\braket{ \pi^a(q) \omega (k') \mid  i  \tilde{J}^{a}_{\mu \bar{\mu}}(k) \mid 0}
& \simeq & f^T_{b_1} \sum_{\lambda} \braket{ \pi^a(q) \omega (k') \mid \bar{b}^a_1(k,\lambda) }
( {\bar{\epsilon}^{(\lambda)^{*}}_{\mu}} k_{\bar{\mu}} -{\bar{\epsilon}^{(\lambda)^{*}}_{\bar{\mu}}} k_{\mu})
\nonumber\\ && \mbox{} 
-f^T_{\rho}  \sum_{\lambda}  \braket{\pi^a(q) \omega (k') \mid \bar{\rho}(k,\lambda) }  
 \frac{\bar{\epsilon}_{\mu \bar{\mu}\alpha\bar{\alpha}} }{2} 
( {\bar{\epsilon}^{(\lambda)^{*}\alpha}} k^{\bar{\alpha}} - {\bar{\epsilon}^{(\lambda)^{*}\bar{\alpha}}} k^{\alpha} ) 
\nonumber\\ && ~\mbox{} 
+ \dots, 
\label{b1decay}
\end{eqnarray}
where $\tilde{J}^{a}_{\mu \bar{\mu}}(k) \equiv -\frac12 \epsilon_{\mu\bar{\mu}\alpha\bar{\alpha}}(k^2-\bar{m}^2)\int d^4x e^{ikx}
\bar{q}(x) T^a \sigma^{\alpha\bar{\alpha} } q(x) $ with $\bar{m}$ being the meson mass in the tensor representation. 
In the soft pion limit, the overlap can be rewritten as
\begin{align}
\braket{ \pi^a(q) \omega (k') \mid i   \tilde{J}^{a}_{\mu \bar{\mu}}(k) \mid 0}
& = \frac{i}{ f_{\pi}} \int d^3x e^{-i {\bm{q}} \cdot {\bm{x}}}  
\braket{ \omega (k') \mid [J^a_{5 0}(x), i\tilde{J}^{a}_{\mu \bar{\mu}}(k) ] \mid 0}
\nonumber\\
&= \frac{i}{  f_{\pi}} \left[ -3i \braket{ \omega (k) \mid  i J^{0}_{\mu\bar{\mu}}(k) \mid 0 } + R_1(q) \right]
\nonumber\\
& \simeq  \frac{3f^T_{\omega}}{ f_{\pi}} \sum_{\lambda} \braket{\omega (k) \mid \bar{\omega}(k,\lambda)}   
( {\bar{\epsilon}^{(\lambda)^{*}}_{\mu}} k_{\bar{\mu}} - {\bar{\epsilon}^{(\lambda)^{*}}_{\bar{\mu}}} k_{\mu}) 
+ \dots,
\label{b1decayo}
\end{align}
where $J^{0}_{\mu \bar{\mu}}(k) \equiv  (k^2-\bar{m}^2) \int d^4x e^{ik x}\bar{q}(x) T^0 \sigma^{\mu\bar{\mu} } q(x)$
and $R_1(q)$ represents $\lvert \bm{q} \rvert$-dependent contributions in the limit of $\lim_{q \rightarrow 0}R_1(q) = 0$. 
Comparing Eqs.~\eqref{b1decay} and \eqref{b1decayo}, the $|\bm{q}|$-independent term can be obtained as
\begin{align}
\lim_{q \rightarrow 0} \braket{ \pi^a(q) \omega (k') \mid \bar{b}^a_1(k,\lambda)} 
\simeq  \frac{3f^T_{\omega}}{f_{\pi} f^T_{b_1}  }  \braket{ \omega (k) \mid \bar{\omega}(k,\lambda) }.
\end{align}
If the $\bar{\omega}$ is identified as the $\omega(782)$ state, one can obtain the strength for the $b_1\rightarrow\pi  \gamma$ 
decay from the overlap for $b_1 \to \pi \omega$ with the VMD hypothesis. 
This process can be written as the following current correlation function in the soft pion limit:
\begin{align}
\braket{ \pi^a(q) \mid J^{0}_{\alpha}(k')i\tilde{J}^{a}_{\mu \bar{\mu}}(k) \mid 0 }
& = \frac{i}{  f_{\pi}} \int d^3 x e^{-i{\bm{q}} \cdot {\bm{x}}} 
\braket{ 0 \mid [J^a_{5 0}(x) , J^{0}_{\alpha} (k) i\tilde{J}^{a}_{\mu \bar{\mu}}(k)] \mid 0 }
\nonumber \\
& = \frac{i}{  f_{\pi}} \bigl [ 
\braket{ 0 \mid  [ Q^a_5, J^{0}_{\alpha}(k) ] i\tilde{J}^{a}_{\mu \bar{\mu}}(k) \mid 0} + 
\braket{ 0 \mid J^{0}_{\alpha}(k) [ Q^a_5,i\tilde{J}^{a}_{\mu \bar{\mu}}(k)  ] \mid 0 } 
\nonumber \\
& \qquad \quad+ 
R_2(q) \bigr]
\nonumber\\
& = \frac{i}{  f_{\pi}} \bigl[   -3 i \braket{ 0 \mid J^{0}_{\alpha}(k)  i J^{0}_{\mu\bar{\mu}}(k)  \mid 0 }  + \dots \bigr]
\nonumber\\
& \simeq \frac{3 f_{\omega}f^T_{\omega}}{f_{\pi}} \sum_{\lambda} 
\braket{ \omega(k,\lambda) \mid \bar{\omega} (k,\lambda)} 
\epsilon^{(\lambda)}_{\alpha}  ( {\bar{\epsilon}^{(\lambda)^{*}}_{\mu}} k_{\bar{\mu}} - {\bar{\epsilon}^{(\lambda)^{*}}_{\bar{\mu}}}
k_{\mu}) + \dots ,
\label{eq:new-eq}
\end{align}
where $J^{0}_{\alpha}(k)\equiv (k^2-m^2)\int d^4x e^{ikx} \bar{q}(x) T^0 \gamma_{\alpha} q(x)$ with $m$ 
being the meson mass in chiral representation and $R_2(q)$ represents the $\vert {\bm{q}} \vert$-dependent contributions
with the condition that $\lim_{q \rightarrow 0}R_2(q) = 0$. 
The overlap $f_{\omega}f^T_{\omega} \braket{ \omega(k,\lambda) \mid \bar{\omega} (k,\lambda) }$ can be calculated
from the time-ordered correlator as
\begin{align}
\Pi^{\bar{b}_1-\pi\omega}(k) = \frac{3}{f_\pi} \int d^4x e^{ikx}
\braket{ \mathcal{T} \left[J^{0}_{\alpha}(x)J^{0}_{\mu\bar{\mu}}(0) \right]}.
\label{b1gamma}
\end{align}
If chiral symmetry is restored, the correlation~\eqref{b1gamma} will vanish because the representation spaces of currents are totally
disconnected. 
Diagrammatical descriptions of this process are depicted in Fig.~\ref{bo}.
Therefore, Eq.~(\ref{eq:new-eq}) describes the decay mechanism of $\bar{b}_1 \to \pi\gamma$ that evades the usual VMD hypothesis.
Then, the interpretation of the $\bar{b}_1$ and $\bar\omega$ is required in connection with physical states, namely, to see
whether and how the tensor representation can describe observed mesons.

\section{Operator product expansion and spectral sum rules}
\label{sec:ope}

Understanding the nature of the $\bar{\omega}$ and $\bar{b}_1$ states requires their relations to the physical 
$\omega(782)$ and $b_1(1235)$ mesons. 
For this purpose, we analyze the spectral sum rules for the corresponding masses of these mesonic states. 
The correlation function can be calculated via OPE in the $k^2 \rightarrow-\infty$ limit as in 
Refs.~\cite{WZ72, SVZ79a, SVZ79b,SVZ79c, RRY85, GRDW87, BB96}.

\subsection{OPE of the correlators}

To interpolate the mesonic state with given quantum numbers we employ the tensor current,
\begin{align}
J^A_{\mu\bar{\mu}}(x)= \bar{q}(x) T^A \sigma_{\mu\bar{\mu}}^{} q(x),
\label{tensorc}
\end{align}
where flavor matrix $T^0$ corresponds to the ($\bar{\omega}$, $\bar{h}_1$)
current and $T^a$ corresponds to ($\bar{\rho}^a$, $\bar{b}_1^a$) current. 
The correlator can be obtained as~\cite{GRDW87, BB96}
\begin{align}
\Pi^{AB}_{\mu \bar{\mu}; \nu \bar{\nu}}(k) &= i \int d^4x e^{ikx}
\braket{ \mathcal{T} [\bar{q}(x) T^A \sigma_{\mu\bar{\mu}}^{} q(x) \,
\bar{q}(0) T^B \sigma_{\nu\bar{\nu}}^{} q(0)] }
\nonumber\\
&=  \Pi^{\rm ope}_{-} (k^2)  P^{(-)}_{\mu \bar{\mu}; \nu \bar{\nu}} + \Pi_{+}^{\rm ope}(k^2)  P^{(+)}_{\mu \bar{\mu}; \nu \bar{\nu}},
\label{tcorr1} \\
\Pi_{\mp}^{\rm ope}(k^2)  &=\frac{\delta^{AB}}{2} \left[\Pi_{\textrm{pert.}}  (k^2)+\Pi_{G^2} (k^2) \pm \Pi_{4q_{(a)}}^{A}  (k^2)  +\Pi_{4q_{(b)}}  (k^2)\right],
\end{align}
where nonzero OPE terms are found as
\begin{align}
\Pi_{\textrm{pert.}} (k^2) & =   \frac{1}{8\pi^2} \left(-k^2\right) \left[ \left(1+\frac{7 \alpha_s}{9 \pi }\right) \ln(-k^2/\mu^2)+\frac{ \alpha_s}{3 \pi } 
\left( \ln(-k^2/\mu^2) \right)^2 \right],
\label{pert}\\
\Pi_{G^2}  (k^2) & =  \frac{1}{24} \left(- \frac{1}{ k^2}\right) \left\langle \frac{\alpha_s}{\pi} G^2 \right\rangle, 
\label{gluon}\\
\Pi_{4q_{(a)}}^{A}  (k^2) & =   -32 \pi \alpha_s \left(-\frac{1 }{k^2}\right)^2 \left( \braket{ \bar{q} T^A \tau^{\bar{a}} q \, \bar{q} T^A \tau^{\bar{a}} q} 
+ \braket{ \bar{q} T^A \tau^{\bar{a}} \gamma_5^{} q \, \bar{q} T^A \tau^{\bar{a}} \gamma_5^{} q} \right),  
\label{4qdisc}\\
\Pi_{4q_{(b)}} (k^2) & = -\frac{8 \pi \alpha_s }{9} \left(-\frac{1 }{k^2}\right)^2 
\braket{ \bar{q} T^0\tau^{\bar{a}} \gamma_\eta^{} q\, \bar{q} T^0 \tau^{\bar{a}}\gamma^\eta q}, \label{4qconn}
\end{align}
with $G_{\rho\sigma}=D_{\rho} A_{\sigma} - D_{\sigma} A_{\rho} $ being the gluon field strength tensor and $\mu=0.5~\textrm{GeV}$ is 
the OPE separation scale where the condensate is normalized.%
\footnote{In Eq.~(\ref{4qdisc}), the index $A$ is not summed over.}
The correlator~\eqref{b1gamma} can be calculated as
\begin{align}
\Pi^{\bar{b}_1-\pi\omega}_{\alpha;\mu\bar{\mu}}(k) &= 
\frac{3}{f_\pi} \int d^4x e^{ikx}  
\braket{ \mathcal{T} [\bar{q}(x) T^0 \gamma_\alpha q(x) \, \bar{q}(0) T^0 \sigma_{\mu\bar{\mu}} q(0) ] }
\nonumber\\
&= \Pi^{\bar{b}_1-\pi \omega}(k^2)P^{(\gamma - \sigma) }_{\alpha; \mu
\bar{\mu}}, \label{b1gamma2}
\end{align}
where
\begin{align}
\Pi^{\bar{b}_1-\pi \omega}(k^2)& = - \frac{3}{f_\pi} \left\{ -\frac{1}{k^2} \braket{ \bar{q} T^0 q} +
\frac{1}{12}  \left(-\frac{1}{k^2}\right)^2  \left\langle \bar{q}
T^0 g \sigma \cdot G q \right\rangle \right\}.
\label{b1gamma3}
\end{align}

\subsection{Spectral sum rules}

The invariants can be weighted to emphasize the ground state contribution as follows:
\begin{align}
\mathcal{W}_M^{\textrm{subt.}} \left[\Pi_{\mp}^{\rm ope} (k^2) \right]
& = \frac{1}{\pi } \int_{0}^{s_0} ds e^{-s /M^2} \mbox{Im} \left[ \Pi_{\mp}^{\rm ope}(s)  \right]
\nonumber\\ 
&= -\frac{1}{16\pi^2} \left[ \left(1+\frac{7 \alpha_s}{9 \pi } \right) (M^2)^2E_1(s_0^{}) +\frac{  \alpha_s}{3 \pi } L(s_0) \right]
\nonumber\\
&\quad \mbox{} \mp \frac{ 16 \pi \alpha_s }{M^2}  \left( \braket{ \bar{q} T^A \tau^{\bar{a}} q \, \bar{q} T^A \tau^{\bar{a}} q}  + 
\braket{ \bar{q} T^A \tau^{\bar{a}} \gamma_5^{} q  \, \bar{q} T^A \tau^{\bar{a}} \gamma_5^{} q } \right) \nonumber\\
&\quad \mbox{} - \frac{4 \pi \alpha_s }{9 M^2}  
\braket{ \bar{q} T^0\tau^{\bar{a}} \gamma_\eta^{} q \, \bar{q} T^0 \tau^{\bar{a}}\gamma^\eta q }  - \frac{1}{48} \left\langle \frac{\alpha_s}{\pi} G^2 \right\rangle, 
\label{btope1}
\end{align}
where `$\mp$' denote parity-odd and parity-even modes and the OPE continuum contribution ($k^2 > s_0^{}$), where the excited states are assumed, 
is subtracted as
\begin{equation}
E_1^{}(s_0^{}) \equiv 1 - e^{-s_0^{}/M^2} \left(1+s_0^{}/M^2 \right).
\label{conth}
\end{equation}
We also define
\begin{equation}
L(s_0) \equiv 2 \int_{0}^{s_0} ds \, e^{-s /M^2} s \ln(s/\mu^2).
\end{equation}
Detailed arguments for the weighting scheme are given in Appendix~\ref{appenixc}.  
The strong coupling $\alpha_s=0.5$ is obtained at the separation scale $\mu = 0.5$~GeV with $\Lambda_{\textrm{QCD}}=0.125$~GeV. 
After Borel weighting, $\alpha_s$ is kept as constant.

The value of the gluon condensate is taken as $\braket{ (\alpha_s/\pi)G^2 }=(0.33~\textrm{GeV})^4$~\cite{SVZ79a, SVZ79b,SVZ79c}. 
By using the factorization hypothesis,
\begin{align}
\langle q _\alpha\bar{q}_\beta
q_\gamma\bar{q}_\delta\rangle_{\rho,I}&\simeq\langle
q_\alpha\bar{q}_\beta \rangle_{\rho,I}\langle
q_\gamma\bar{q}_\delta \rangle_{\rho,I}- \langle
q_\alpha\bar{q}_\delta \rangle_{\rho,I} \langle
q_\gamma\bar{q}_\beta \rangle_{\rho,I},
\end{align}
where the color indices are omitted, the four-quark condensates are estimated in factorized forms as
\begin{align}
\braket{ \bar{q} T^A \tau^{\bar{a}} q \,  \bar{q} T^A \tau^{\bar{a}} q } \rightarrow 
- \frac{\alpha_{A}^{}}{18}  \braket{ \bar{q} T^0  q }_{\textrm vac}^2,
\label{factcond1} \\
\braket{ \bar{q} T^A \tau^{\bar{a}} \gamma_5^{} q \, \bar{q} T^A \tau^{\bar{a}} \gamma_5^{} q } \rightarrow 
-\frac{\beta_{A}}{18} \braket{ \bar{q} T^0  q  }_{\textrm vac}^2,
\label{factcond2}\\
\braket{\bar{q} T^0\tau^{\bar{a}} \gamma_\eta^{} q \, \bar{q} T^0 \tau^{\bar{a}}\gamma^\eta q }
\rightarrow -\frac{2\gamma}{9}  \braket{ \bar{q} T^0  q }_{\textrm vac}^2,
\label{factcond3}
\end{align}
where $\braket{\bar{q} T^0 q }_{\textrm{vac}}$ is obtained from the Gellmann-Oakes-Renner relation~\cite{GOR68},
\begin{align}
2m_q \braket{ \bar{q} T^0 q }_{\textrm{vac}}=-m^2_\pi f^2_\pi,
\end{align}
with $m_\pi=138$~MeV and $f_\pi=93$~MeV. 
This leads to $\braket{ \bar{q} T^0 q }_{\textrm{vac}} \simeq -(254~\mbox{MeV})^3$ for $m_q=5$~MeV. 
The parameters in Eqs.~(\ref{factcond1})-(\ref{factcond3}) have the same value, namely, $\alpha_{A}^{}=\beta_{A}^{}=\gamma=1$, 
in the usual factorization hypothesis. 
Hereafter, each parameter set with explicit isospin dependence, i.e., $\alpha_{0}^{}$, $\beta_{0}^{}$, 
$\alpha_{a}^{}$, and $\beta_{a}^{}$ with $a=1,2,3$, represents the factorization of the isoscalar and isovector 
four-quark condensates, respectively. 
The anomalous running of $\braket{ \bar{q} T^0 q }_{\textrm{vac}}^2 $ is neglected because the factorization scheme is 
used only for estimating the four-quark condensate scale. 
Various scales of $\alpha$, $\beta$, and $\gamma$ are assumed in the present work to include variations coming from 
the running of the condensates.
As can be seen from  Eqs.~(\ref{factcond1})-(\ref{factcond3}), the four quark condensates can be written 
as a product of two quark antiquark pair.  
Depending on the color and flavor matrices appearing in the quark antiquark pair, the factorization parameters 
$\alpha$, $\beta$, $\gamma$ can take different values.  
This is so because, when there is an isospin operator in the quark antiquark pair, the quark-disconnected diagrams 
vanish identically. 
Furthermore, even for isosinglet quark antiquark pair, depending on the chiral and color structure, the quark-disconnected 
diagrams are expected to have different contributions.  
Therefore, in the present work, we choose the parameter set taking into account the following three conditions 
depending on the color and isospin structure of the quark antiquark pair:
(a)~possible correlation types that depend on chiral symmetry breaking, 
(b)~nontrivial topological contribution from color gauge group, 
and (c)~stability of the Borel curves. 
For isovector quark-antiquark pair, the usual factorization provides quark-connected-like condensates. 
The condensates are given in the local limit of the pair of gauge equivalent nonlocal bilinear~\cite{CD81}. 
The colored pieces are part of the Dirac eigenmodes in the nonlocal gauge equivalent structure of the mesonic current. 
So the isoscalar condensates are considered to have nonzero quark-disconnected-like contribution if the colored
spinless bilinear appears in pair to produce the totally color singlet configuration.  
Considering the possibility of this contribution, which does not appear in the usual factorization scheme, 
the quark-disconnected-like and quark-connected-like contributions are encoded in the isoscalar and 
isovector parameters, respectively. 
This typically leads to
$\vert \alpha_{0}\vert < \vert \alpha_{a}\vert$ and $\vert \beta_{0}\vert <\vert \beta_{a}\vert$.
Also, if one considers topologically nontrivial gauge contributions, the spinless but party-odd combination of quark bilinear can have nonzero
correlation in the same way as argued in Refs.~\cite{LS92b, EHS96a}. 
For the condensate composed with pseudoscalar bilinear in Eq.~\eqref{factcond3}, the nontrivial contribution may
appear as an alternating series in winding number of the color gauge. 
%So the magnitude of the parameter sets $\vert \beta_{A}^{} \vert$ are taken to be smaller than $\vert \alpha_{A}^{} \vert$. 
Sum rule analysis with the parameters taken according to these criteria are found to produce stable Borel curves.

%%%%%% FIG 2
\begin{figure}
\centering
\includegraphics[height=5.9cm]{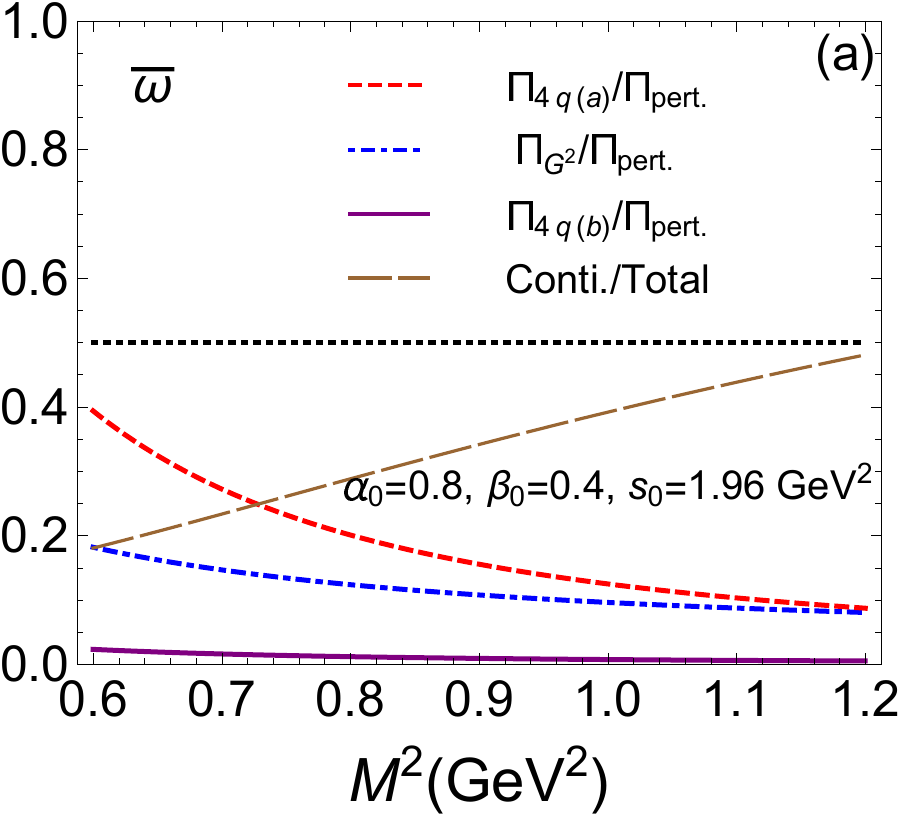} \qquad
\includegraphics[height=5.9cm]{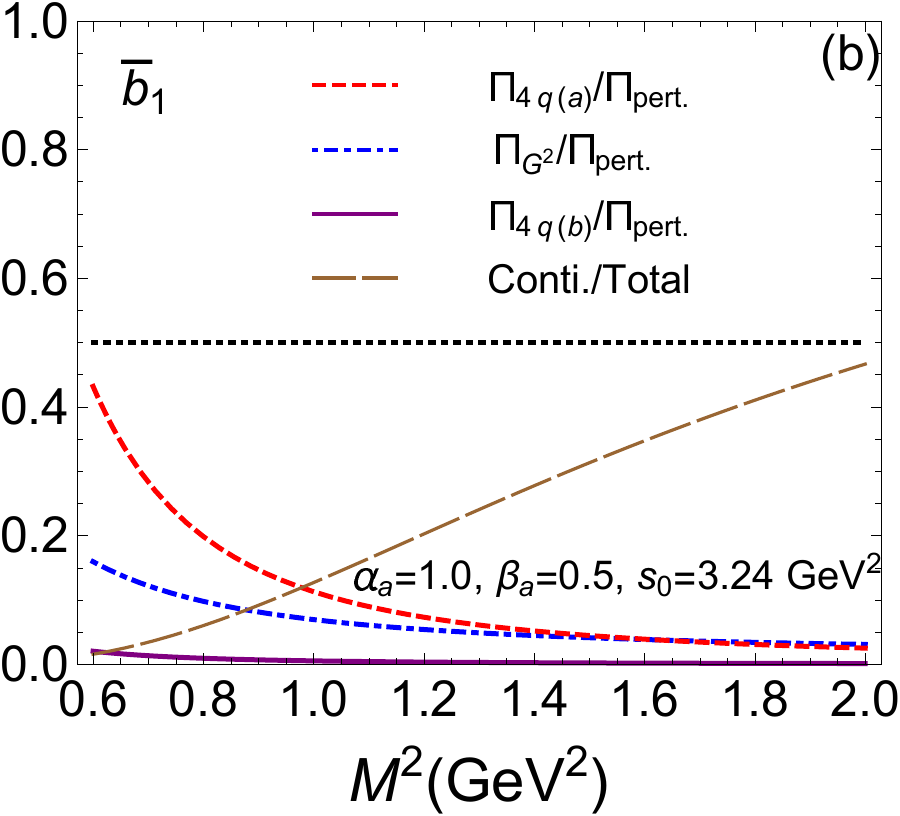}
\caption{Spectral structure of the weighted invariant (a) for the $\bar{\omega}$ and (b) for the $\bar{b}_1$. 
As a reference, black dotted lines are given to denote 50\%.}\label{bw}
\end{figure}

We first take ($\alpha_0^{}=0.8$, $\alpha_a^{}=1.0$) as the reference parameter set, which is similar to the usual factorization.%
\footnote{We leave short technical remarks for the sum rules. 
Some parameter sets with ($\alpha_A^{} <0$, $\beta_A^{}<0$) are found to give a mass of the parity-odd mode close to $\sim 800~\textrm{MeV}$. 
However, these parameter sets lead to unstable Borel curves and controversial results for the parity-even mode sum rules. 
Therefore, we do not adopt these parameter sets in our study. More detailed sum rules analyses for $\bar{\omega}$ and $\bar{b}_1$ states 
with ($a_A^{}<0$, $b_A^{}<0$) are reported in Appendix~\ref{appenixc}. }
To ensure that the ground state contribution becomes dominant and the OPEs are convergent, the Borel windows for the $\bar{\omega}$ 
and $\bar{b}_1$ sum rules are chosen to be in the range of $0.6~\textrm{GeV}^2 < M^2 < 1.2~\textrm{GeV}^2$ and 
$0.6~\textrm{GeV}^2 < M^2 < 2.0~\textrm{GeV}^2$, respectively. 
Plotted in Fig.~\ref{bw} are spectral structures of the invariants.

If the ground state is the one-particle mesonic state as explained in Sec.~\ref{sec:tensor}, 
the phenomenological structure can be written as
\begin{align}
\Pi_{\mp}^{\textrm{ph.pole}}(k^2) & = \frac{ {(f^{T}_{\mp})}^2 k^2}{k^2-m_{\mp}^2+i\epsilon}
\nonumber \\
& =\mathcal{P}\frac{ {(f^{T}_{\mp})}^2 k^2}{(k^2-m_{\mp}^2)}-i \pi {(f^{T}_{\mp})}^2 k^2\delta(k^2-m_{\mp}^2).
\end{align}
However, it is also possible that the tensor current~\eqref{tensorc} can couple to some kinds of hybrid states, which are composed of two particles, 
if the mass scale of the hybrid is compatible with the one-particle mesonic state. 
The hybrid state containing the quantum number of the $\omega$ meson can be inferred from the anomalous interaction that is written as~\cite{Meissner88}
\begin{align}
\mathcal{L}^{\epsilon1}_{\omega \pi\rho}&=\frac{g_{\omega
\pi\rho}}{2}\epsilon^{\mu \bar{\mu}\alpha \bar{\alpha}}
\omega_{\mu \bar{\mu}}
\partial_{\alpha}\pi^a \rho_{\bar{\alpha}}^{a}=-g_{\omega
\pi\rho}\epsilon^{\mu \bar{\mu}\alpha \bar{\alpha}}\omega_{\bar{\mu}}
\partial_{\alpha}\pi^a \partial_{\mu}\rho_{\bar{\alpha}}^{a}, \label{omegaac}
\end{align}
where $\omega_{\mu \bar{\mu}}$ is the $\omega$ field strength tensor. 
As the $\omega$ and $\bar{\omega}$ states share the same quantum numbers $[0^{-}(1^{--})]$,  the phenomenological interpolating current
for the $\bar{\omega}$ state can be written as
\begin{align}
J_{\mu \bar{\mu}}^{\bar{\omega}}(x)& \equiv  \epsilon_{\mu \bar{\mu} \alpha \bar{\alpha}} \textrm{Tr}\left[
\partial^{\alpha}\pi(x)  \rho ^{\bar{\alpha}}(x)\right],\label{phencurrent}
\end{align}
where the trace is summed over flavor indices. 
The corresponding phenomenological structure of $\Pi_{(\mp)}^{\bar{\omega}}(k^2)$ can
be obtained from the correlator of $J_{\mu \bar{\mu}}^{\bar{\omega}}$ as
\begin{align}
\Pi^{\bar{\omega} }_{\mu \bar{\mu}; \nu \bar{\nu}}(k) &=
i \int d^4x e^{ikx} \left\langle \mathcal{T}
\left[\epsilon_{\mu
\bar{\mu}\alpha \bar{\alpha}} \textrm{Tr} \big[ \partial^{\alpha} \pi(x) \rho^{\bar{\alpha}} (x) \big]
\epsilon_{\nu \bar{\nu}\beta\bar{\beta}} 
\textrm{Tr} \big[\partial^{\beta} \pi(0) \rho^{\bar{\beta}}(0)  \big]\right] \right\rangle
\nonumber\\
&= \frac{3  i}{4}  \epsilon_{\mu
\bar{\mu}\alpha \bar{\alpha}}  \epsilon_{\nu \bar{\nu}\beta
\bar{\beta}} \int \frac{d^4p}{(2\pi)^4} \frac{(p+k)^{\alpha}(p+k)^{\beta}}{(p+k)^2-m_{\pi}^2} \frac{1}{p^2-m_{\rho}^2} \left( g^{\bar{\alpha}\bar{\beta}}-\frac{p^{\bar{\alpha}}p^{\bar{\beta}}}{m_{\rho}^2} \right) \nonumber\\
&=  \Pi^{\bar{\omega} }_{-}
(k^2) P^{(-)}_{\mu \bar{\mu}; \nu \bar{\nu}}+\Pi_{+}  (k^2) P^{(+)}_{\mu \bar{\mu}; \nu \bar{\nu}},
\label{phencorr}
\end{align}
where $m_\rho^{}$ is the $\rho$ meson mass and the invariant function for each parity mode can be calculated as
\begin{eqnarray}
\Pi^{\bar{\omega} }_{-} (k^2) &=& \frac{3}{4(4\pi)^2}\int^1_0 dx \left(-x(1-x)k^2
+(1-x)m_{\rho}^2\right)\ln{\left(-x(1-x)k^2
+(1-x)m_{\rho}^2\right)}
\nonumber \\ && \mbox{}
+R_{-}(k^2),
\label{phencorrn}\\
\Pi_{+} (k^2) &=& \frac{3 }{4(4\pi)^2}\int^1_0 dx
\left[-\left(1-x\right)^2k^2+\left(x\left(1-\frac{x}{2}\right)\frac{k^2}{m^2}-1\right)\left(-x(1-x)k^2
+(1-x)m_{}^2\right)\right]
\nonumber\\ &&\qquad\qquad\qquad\quad \mbox{}
\times \ln \left[-x(1-x)k^2 +(1-x)m_{}^2\right]
+R_{+}(k^2),\label{phencorrp}
\end{eqnarray}
where $R_{\mp}(k^2)$ is a regular polynomial and the $m_{\pi} \rightarrow 0$ limit has been taken. 
One can verify that the imaginary part appears in the range of $m_{\rho}^2/k^2 < x<1$. 
The weighted $\Pi^{\bar{\omega} }_{-}(k^2)$ can be obtained as
\begin{eqnarray}
\mathcal{W}_{M}^{\textrm{subt.}}
\left[\Pi^{\bar{\omega} }_{-}(k^2)\right] &=&
\frac{1}{\pi } \int_{m_{\rho}^2}^{s_0^{}} ds\, e^{-s /M^2} \, \textrm{Im} \left[ \Pi^{\bar{\omega} }_{-} (s)\right]
\nonumber\\
&=& 
\frac{3}{64\pi^2}\int^{s_0^{}}_{m_{\rho}^2} ds\, e^{-s/M^2} \left( -\frac{ s}{6}+\frac{m^2_{\rho}}{2}-\frac{1}{2}\frac{(m^2_{\rho})^2}{s} 
+\frac{1}{6}\frac{(m^2_{\rho})^3}{s^2}  \right)
\nonumber\\
&=& \frac{3}{64\pi^2}\Bigg\{ \frac{M^2}{6}\left( s_0^{} e^{-s_0^{}/M^2} -m^2_{\rho}e^{-m^2_{\rho}/M^2} \right)
%\nonumber\\&&\qquad
 + \frac{(M^2)^2}{6} \left(  e^{-s_0^{}/M^2} - e^{-m^2_{\rho}/M^2} \right)
 \nonumber\\
&&\qquad \quad \mbox{}
-\frac{ m^2_{\rho}}{2}M^2\left( e^{-s_0^{}/M^2} - e^{-m^2_{\rho}/M^2} \right)
\nonumber\\&&\qquad \quad \mbox{}
-\frac{(m^2_{\rho})^2}{2} \left[ \Gamma\left(0,m^2_{\rho}/M^2\right)-\Gamma\left(0, s_0^{}/M^2\right) \right] 
\nonumber\\
&&\qquad \quad \mbox{}
- \frac{(m^2_{\rho})^3}{6} \bigg[ \frac{e^{-s_0/M^2} }{s_0}
-\frac{e^{-m^2_{\rho}/M^2}}{m^2_{\rho}} 
\nonumber\\
&&\qquad\qquad \qquad \qquad \mbox{}
+\frac{1}{M^2} \left\{
\Gamma\left(0,m^2_{\rho}/M^2\right)-\Gamma\left(0,s_0/M^2\right)
\right\}\bigg]\Biggr\},
\end{eqnarray}
where the continuum threshold $s_0^{}$ is taken to be the same as that assigned in 
$ \mathcal{W}_M^{\textrm{subt.}} [\Pi_{-}^{\textrm{ope}}(k^2)]$, 
and $\Gamma\left(t, x \right)=\int_x^{\infty}ds \, s^{t-1} \, e^{-s}$ is the incomplete Gamma function.

Similarly, the loop-like phenomenological current for $\bar{b}_1$ state can be defined as
\begin{align}
J_{\mu \bar{\mu}}^{\bar{b}_1a}(x)& \equiv  \frac{\sqrt{3}}{2} \epsilon_{\mu \bar{\mu} \alpha \bar{\alpha}}^{}  
\partial^{\alpha}\pi^{a} (x)  \omega ^{\bar{\alpha}}(x).\label{phencurrent2}
\end{align}
The corresponding invariant can be obtained from the parity-even mode of Eq.~\eqref{phencorrp} 
by replacing $m$ by $m_{\omega}$. 
The imaginary part appears in the range of $m_{\omega}^2/k^2 < x<1$ and the weighted 
$\Pi^{\bar{b}_1}_{+}(k^2)$ is analogously obtained as
\begin{eqnarray}
&& \mathcal{W}_{M}^{\textrm{subt.}} \left[\Pi^{\bar{b}_1}_{+}(k^2) \right]
= \frac{1}{\pi } \int_{m_{\omega}^2}^{s_0^{}} ds\, e^{-s /M^2} \, \textrm{Im} \left[\Pi^{\bar{b}_1}_{+}(s)\right]
\nonumber\\
&= &
\frac{3}{64\pi^2} \int^{s_0^{}}_{m_{\omega}^2} ds \, e^{-s/M^2}  
\left (-\frac{ 1}{6}\frac{ s^2}{m_{\omega}^2} -\frac{ s}{6}+ \frac{m^2_{\omega}}{4 }+ \frac{1}{3} \frac{(m_{\omega}^2)^2}{s } -\frac{1}{4}\frac{(m_{\omega}^2)^3}{s^2} +\frac{1}{10}\frac{(m_{\omega}^2)^4}{s^3} \right)
\nonumber\\ &= &
\frac{3}{64\pi^2} \Bigg\{ 
\frac16 \frac{M^2}{m_{\omega}^2} \left( s_0^2 e^{-s_0^{}/M^2} -(m_{\omega}^2)^2e^{-m_{\omega}^2/M^2} \right)
%\nonumber\\ &&\qquad \mbox{}
+ \frac{1}{3}\frac{(M^2)^2}{m_{\omega}^2}  \left( s_0^{} e^{-s_0^{}/M^2} -m_{\omega}^2e^{-m_{\omega}^2/M^2} \right)
\nonumber\\ &&\qquad \quad \mbox{}
 + \frac{1}{3}\frac{(M^2)^3}{m_{\omega}^2} \left(  e^{-s_0^{}/M^2} - e^{-m_{\omega}^2/M^2} \right) 
+ \frac{M^2}{6}  \left( s_0^{} e^{-s_0^{}/M^2} -m_{\omega}^2e^{-m_{\omega}^2/M^2} \right) 
 \nonumber\\ &&\qquad \quad \mbox{}
+ \frac{(M^2)^2}{6} \left(  e^{-s_0^{}/M^2} - e^{-m_{\omega}^2/M^2} \right)
-\frac{ m_{\omega}^2}{4} \left[ M^2\left( e^{-s_0/M^2} - e^{-m_{\omega}^2/M^2} \right)\right]
\nonumber\\ &&\qquad \quad \mbox{}
+\frac{(m_{\omega}^2)^2}{3}\left[ \Gamma\left(0,m_{\omega}^2/M^2\right)-\Gamma\left(0,s_0/M^2\right) \right] 
\nonumber\\ &&\qquad \quad \mbox{}
+ \frac{(m_{\omega}^2)^3}{4} \left[
 \frac{ e^{-s_0/M^2}}{s_0} - \frac{e^{-m_{\omega}^2/M^2}}{m_{\omega}^2}
 + \frac{1}{M^2} \left[
\Gamma\left(0,m_{\omega}^2/M^2\right)-\Gamma\left(0,s_0/M^2\right)
\right]\right]
\nonumber\\ &&\qquad \quad \mbox{}
- \frac{(m_{\omega}^2)^4}{20} \Bigg[
 \frac{e^{-s_0/M^2} }{s^2_0} -\frac{e^{-m_{\omega}^2/M^2}}{(m_{\omega}^2)^2}
- \frac{1}{M^2}\left(\frac{e^{-s_0/M^2}}{s_0}
 -\frac{e^{-m_{\omega}^2/M^2}}{m_{\omega}^2}
\right)
\nonumber\\ &&\qquad\qquad\qquad  \qquad \mbox{}
-\left(\frac{1}{M^2}\right)^2 \left(
\Gamma\left(0,m_{\omega}^2/M^2\right)-\Gamma\left(0,s_0/M^2\right)
\right) \Bigg]
\Bigg\}.
\end{eqnarray}

%%%%%% FIG 3
\begin{figure}[t]
\centering
\includegraphics[height=5.9cm]{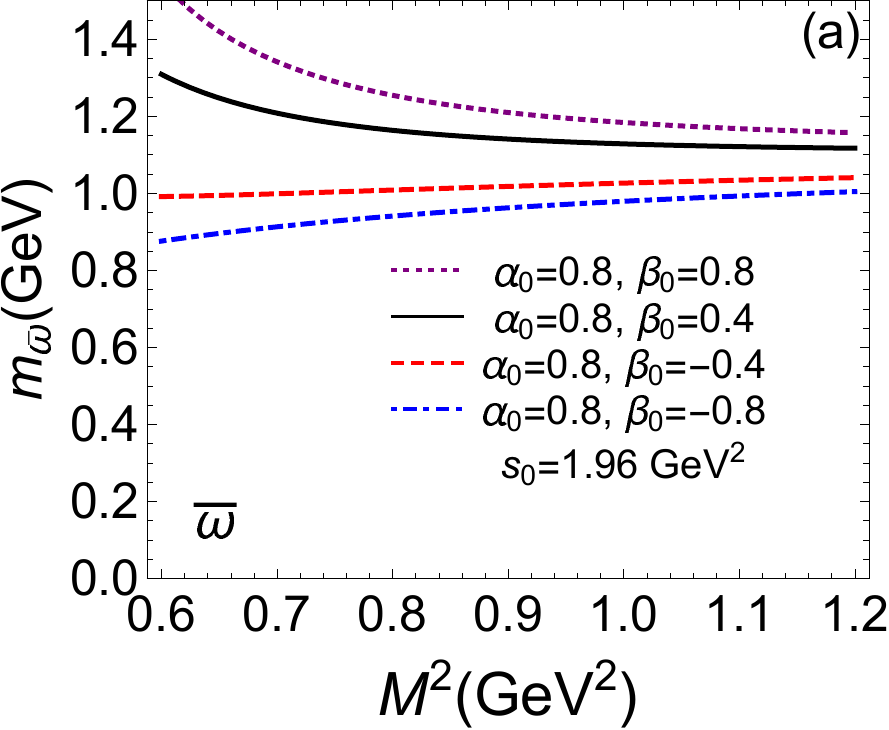} \quad
\includegraphics[height=5.9cm]{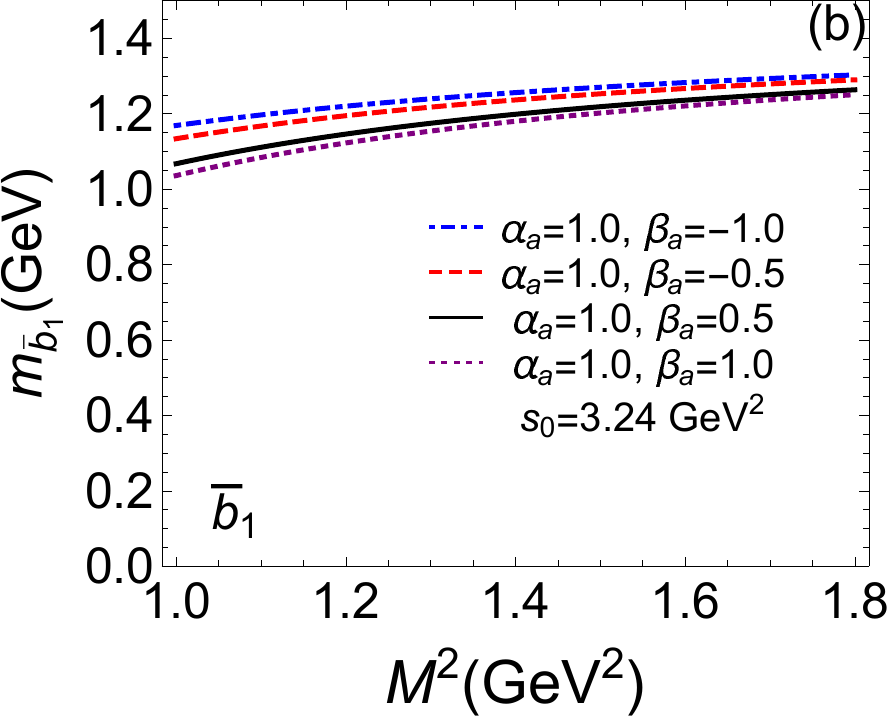}
\includegraphics[height=5.9cm]{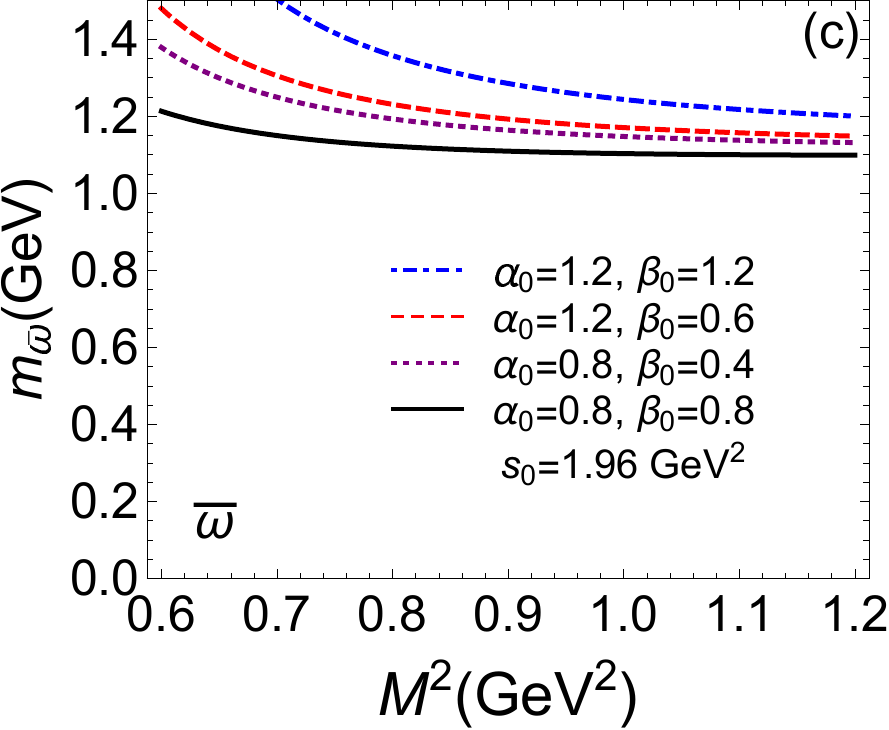} \quad
\includegraphics[height=5.9cm]{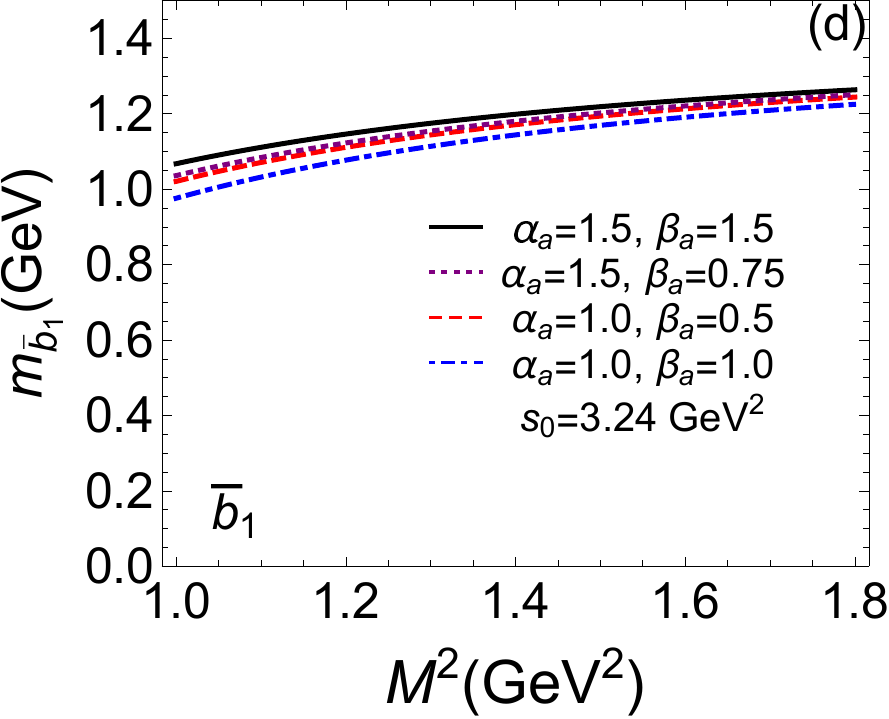}
\caption{Mass sum rules for spin-1 meson states interpolated by the tensor current ($c=1$). 
(a,c) isoscalar parity-odd channel ($\bar{\omega}$) with $s_0^{}=1.96~\si{GeV}^2$ and 
(b,d) isovector parity-even channel ($\bar{b}_1$) with $s_0^{}=3.24~\si{GeV}^2$. 
}
\label{tcsr}
\end{figure}

Then the sum rule for the ground states interpolated by the tensor current reads
\begin{equation}
\mathcal{W}_M^{\textrm{subt.}} \left[\Pi_{\mp}^{\rm ope}(k^2) \right]
= -{(f^{T}_{\mp})}^2 m^2_{\mp}e^{-m^2_{\mp}/M^2}
+{(c^{\bar{\textrm{hy.}}}_{\mp}) }^2\mathcal{W}_{M}^{\textrm{subt.}} \left[\Pi^{\bar{\textrm{hy.}}}_{\mp}(k^2) \right],\label{sr}
\end{equation}
where ${c^{\bar{\textrm{hy.}}}_{\mp} }$ represents the coupling between the tensor current~\eqref{tensorc} and the hybrid state. 
Equation~\eqref{sr} leads to the hybrid-subtracted mass sum rules as
\begin{align}
m_{\mp}=M^2 \left( \frac{(\partial/\partial M^2)\left(
\mathcal{W}_M^{\textrm{subt.}} \left[\Pi_{\mp}^{\rm ope}(k^2)\right]-{(c^{\bar{\textrm{hy.}}}_{\mp} )}^2\mathcal{W}_{M}^{\textrm{subt.}} 
\left[\Pi^{\bar{\textrm{hy.}}}_{\mp}(k^2) \right] \right)} {\mathcal{W}_M^{\textrm{subt.}}
\left[\Pi_{\mp}^{\rm ope}(k^2) \right]-{(c^{\bar{\textrm{hy.}}}_{\mp}) }^2 \mathcal{W}_{M}^{\textrm{subt.}} \left[\Pi^{\bar{\textrm{hy.}}
}_{\mp}(k^2) \right]}\label{mass1}\right)^{\frac{1}{2}}.
\end{align}
By assuming only the one-particle meson state for the ground state, the obtained masses of the $\bar{\omega}$ and $\bar{b}_1$ modes are
plotted respectively in Figs.~\ref{tcsr}(a) and \ref{tcsr}(b).  
Depending on the values of $\alpha_0^{}$ and $\beta_0^{}$, the mass of the isoscalar parity-odd mode is found to vary from $1.0~\si{GeV}$
to $1.2~\si{GeV}$, which is much higher than the physical $\omega(782)$ mass. 
On the other hand, the mass of the isovector parity-even mode shows week dependence on the parameter set, 
and is in good agreement with the physical $b_1(1235)$ mass when $(\alpha_a^{}=1$, $\beta_a^{} = 0.5)$ is employed. 
Presented in Figs.~\ref{tcsr}(c) and \ref{tcsr}(d) are the mass curves for the cases where these parameters are larger than 1.
In this case, the mass curves of the isoscalar parity-odd mode become larger than $1.2~\si{GeV}$ and unstable. 
On the other hand, the mass curves of isovector parity-even mode are not drastically changed.
Because of the large mass difference between the $\omega(782)$ and $\bar{\omega}$, it is unnatural to identify the $\bar{\omega}$ state 
as the physical $\omega(782)$.

%%%%%% FIG 4
\begin{figure}
\centering
\includegraphics[height=5.9cm]{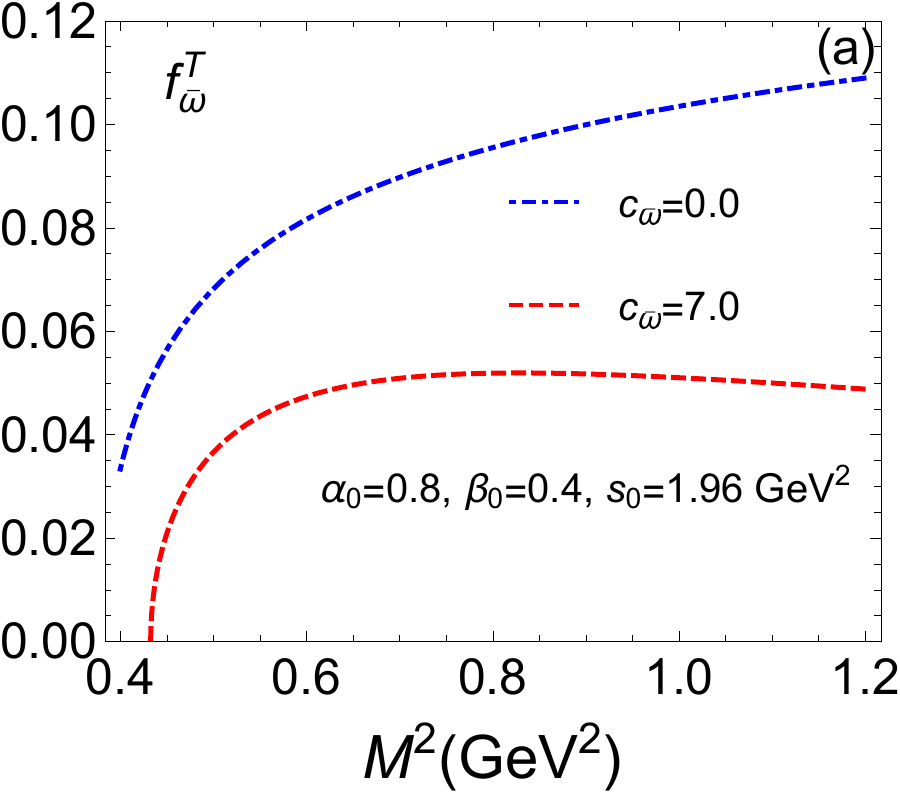} \qquad
\includegraphics[height=5.9cm]{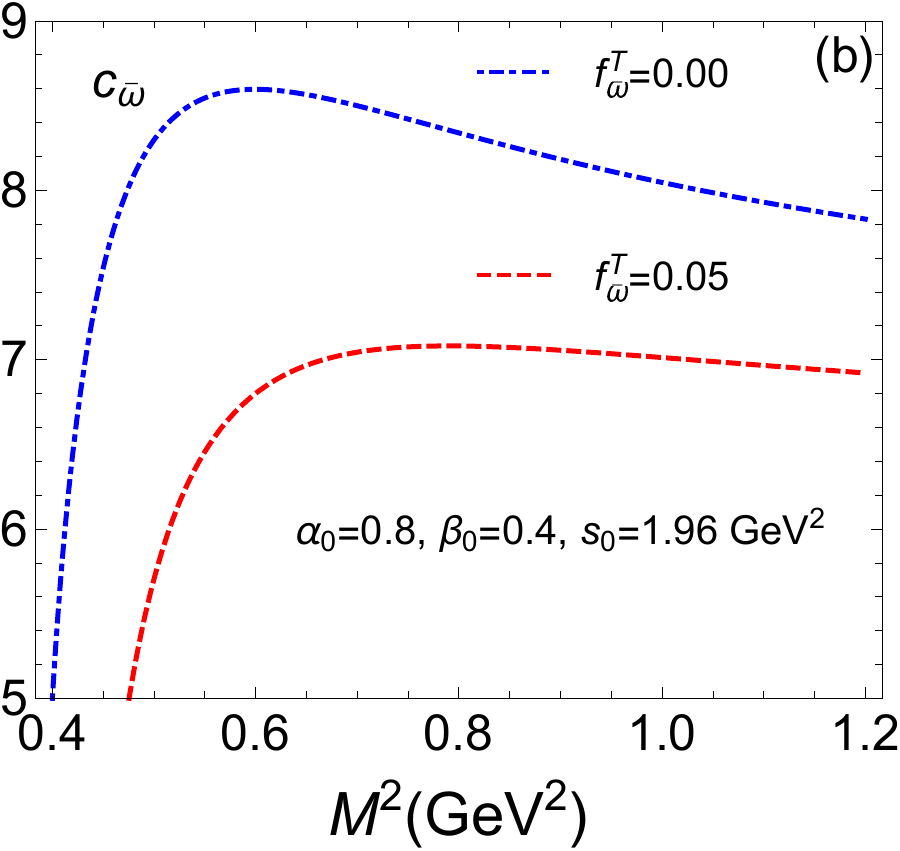}
\caption{Borel curve (a) for the meson pole residue $f^{T}_{\bar{\omega} }$ and 
(b) for the tensor current-hybrid coupling $ c^{\bar{\textrm{hy.}}}_{\bar{\omega} }$.} 
\label{coupling1}
\end{figure}

Considering the mass range determined by the Borel curve in Fig.~\ref{tcsr}(a) and the quantum numbers $[0^{-}(1^{--})]$, one
may imagine that the adopted current couples to excited $\omega$ resonance states. 
However, such possibility is unlikely as the mass of the lowest excited state $\omega(1420)$~\cite{PDG16} 
is much larger than the mass range obtained in the sum rules analysis for the $\bar\omega$. 
Therefore, we explore the possibility that the employed current couples to a hybrid state composed of two particles.  
If the tensor current~\eqref{tensorc} strongly couples to the one-particle mesonic state (hybrid state), 
the sum rules for the one-particle meson pole residue $f^{T}_{\mp}$ (the tensor current-hybrid 
coupling $c^{\bar{\textrm{hy.}}}_{\mp}$) should provide a stable behavior in the Borel window. 
The sum rules for each coupling obtained from Eq.~\eqref{sr} read
\begin{align}
{f^{T}_{\mp}}  =&
\left\{ \frac{1}{m^2_{\mp}}e^{m^2_{\mp}/M^2}\left(-\mathcal{W}_M^{\textrm{subt.}} \left[\Pi_{\mp}^{\rm ope}(k^2) \right]
+ {(c^{\bar{\textrm{hy.}}}_{\mp}) }^2 \mathcal{W}_{M}^{\textrm{subt.}} \left[\Pi^{\bar{\textrm{hy.}}
}_{\mp}(k^2) \right] \right)\right\}^{\frac{1}{2}},\\
{c^{\bar{\textrm{hy.}}}_{\mp} }=&
\left\{ \left(\mathcal{W}_M^{\textrm{subt.}} \left[\Pi_{\mp}^{\rm ope}(k^2) \right]
+{(f^{T}_{\mp})}^2 m^2_{\mp}e^{-m^2_{\mp}/M^2}\right)\bigg/ \mathcal{W}_{M}^{\textrm{subt.}} \left[\Pi^{\bar{\textrm{hy.}}
}_{\mp}(k^2) \right]\right\}^{\frac{1}{2}},
\end{align}
where  $m^2_{\mp}$, $f^{T}_{\mp}$, and  $c^{\bar{\textrm{hy.}}}_{\mp}$ are input parameters. 
If the one-particle pole is dominant in the ground state, the Borel curve for $f^{T}_{\bar{\omega}}$ will be stable. 
However, as can be seen in Fig.~\ref{coupling1}(a), the Borel curve is not stable, suggesting that something is missing or the assumption is not correct.
On the other hand, $c^{\bar{\textrm{hy.}}}_{\bar{\omega}}$ provides relatively stable behavior in the proper Borel window 
even with the no-pole scenario ($f^{T}_{\bar{\omega}}=0$) and provides a plateau value $c^{\bar{\textrm{hy.}}}_{\bar{\omega}}\simeq7.0$ 
with $f^{T}_{\bar{\omega}}=0.05$ as shown in Fig.~\ref{coupling1}(b). 
In these optimized curves, $f^{T}_{\bar{\omega}}$ reduces to $50\%$ of its original scale when $c^{\bar{\textrm{hy.}}}_{\bar{\omega}}$ is varied from
$0$ to $7.0$, while $c^{\bar{\textrm{hy.}}}_{\bar{\omega}}$ reduces only to $85\%$ of its original scale when $f^{T}_{\bar{\omega}}$ varies from $0$ to $0.05$. 
This leads to the conclusion that the ground state information in the tensor current correlator is dominated by the hybrid state.

%%%%%% FIG 5
\begin{figure}
\centering
\includegraphics[height=5.9cm]{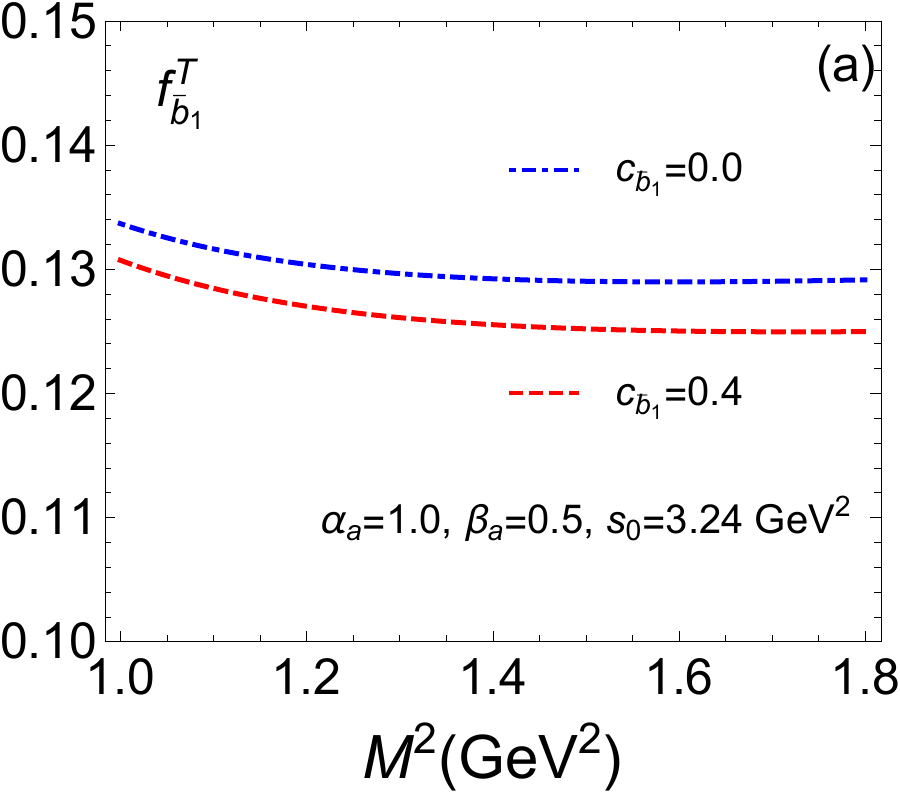} \qquad
\includegraphics[height=5.9cm]{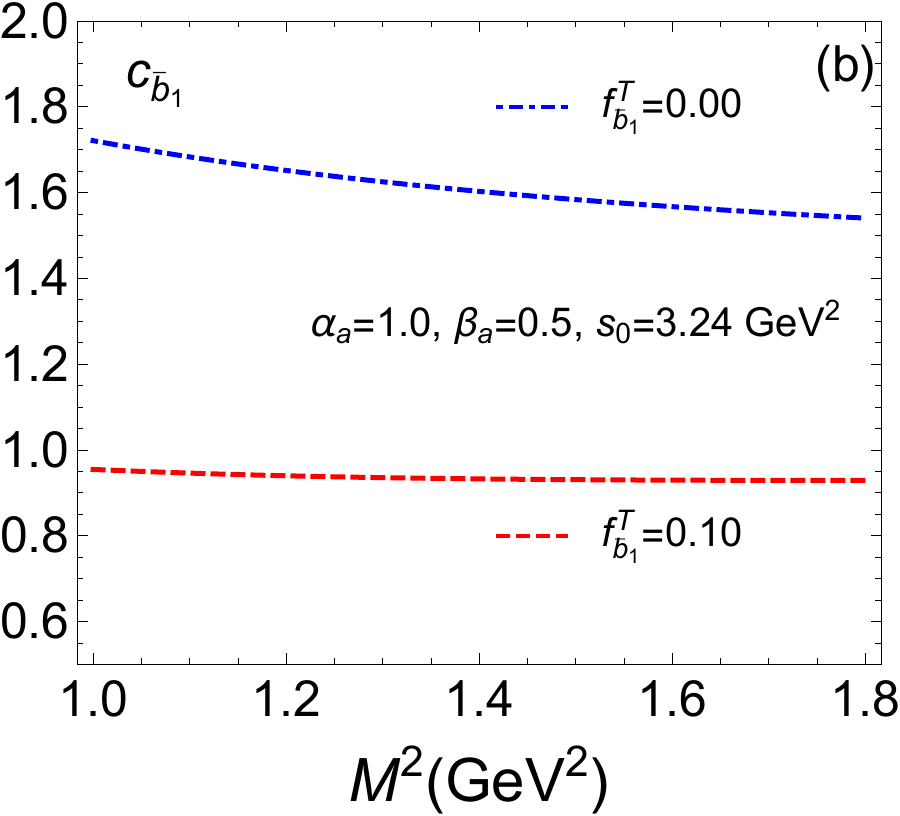} 
\caption{Borel curve (a) for the meson pole residue $f^{T}_{{\bar{b}}_1 }$ and  (b) for the tensor current-hybrid coupling $ c^{\bar{\textrm{hy.}}}_{{\bar{b}}_1 }$.} \label{couplingb}
\end{figure}

The $\bar{b}_1$ state can be analyzed in a similar way. 
As shown in Fig.~\ref{couplingb}, the Borel curves for the couplings in the $\bar{b}_1$ state shows the opposite tendency 
in comparison with the $\bar{\omega}$ analyses.  
Namely, the computed $f^{T}_{{\bar{b}}_1 }$ values show  plataeu-like behavior even in the no-continuum condition 
($c^{\bar{\textrm{hy.}}}_{{\bar{b}}_1 }=0.0$) and reduces only to $90\%$ of its original scale when 
$c^{\bar{\textrm{hy.}}}_{{\bar{b}}_1 }=0.5$ is assigned.  
It is also found that $c^{\bar{\textrm{hy.}}}_{{\bar{b}}_1 }$ strongly depends on $f^{T}_{{\bar{b}}_1 }$ variation and shows singular 
behavior when $f^{T}_{{\bar{b}}_1 }>0.11$ is assigned.  
Therefore, the ${\bar{b}}_1$ state is identified as the physical $b_1(1235)$ meson state considering the mass scale shown in Fig.~\ref{tcsr}(b).

 %%%%%% FIG 6
\begin{figure}
\centering
\includegraphics[height=5.9cm]{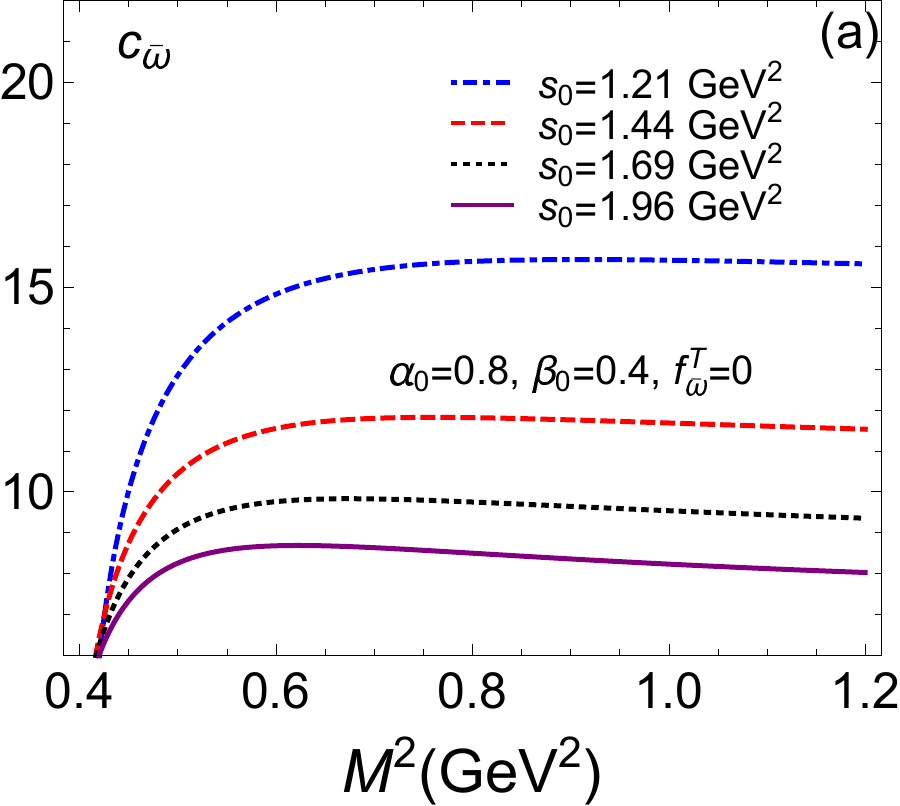} \qquad
\includegraphics[height=5.97cm]{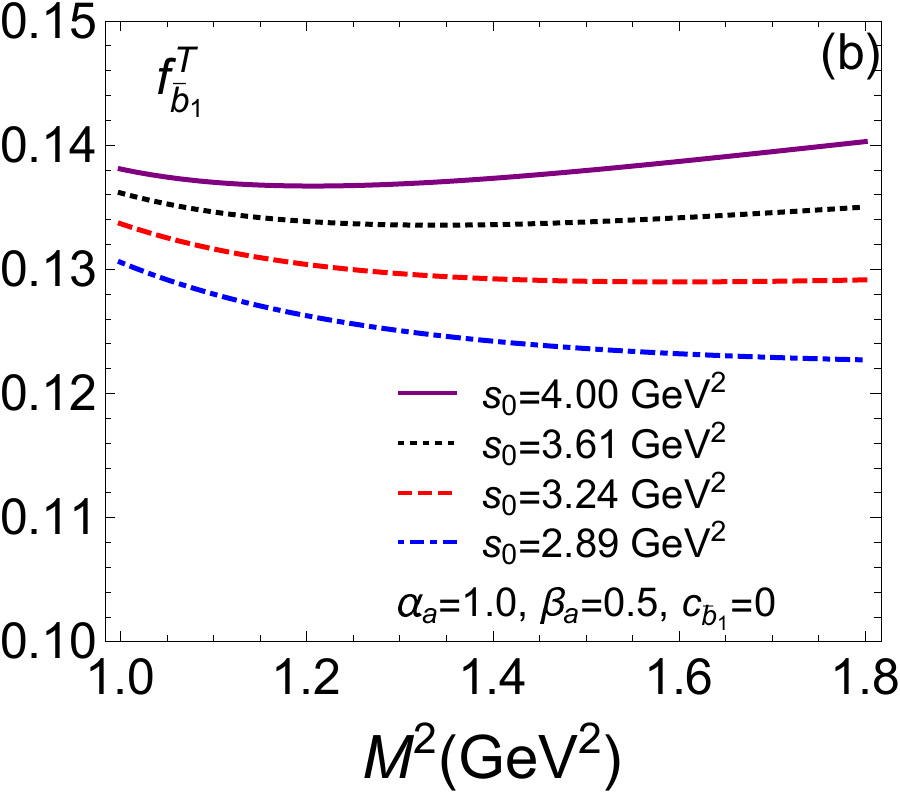} 
\caption{Coupling constants sum rules for (a) the tensor current-hybrid coupling $ c^{\bar{\textrm{hy.}}}_{{\bar{\omega}} }$  
($f^{T}_{\bar{\omega}}=0.00$) and (b) the meson pole residue $f^{T}_{{\bar{b}}_1 }$ ($ c^{\bar{\textrm{hy.}}}_{{\bar{b}}_1 }=0.0$) 
in various continuum thresholds.}
 \label{couplingc}
\end{figure}

By assuming only the $\pi$-$\rho$ hybrid (i.e., the $\bar\omega$) or $b_1(1235)$ state for the ground state, 
the corresponding coupling sum rules are plotted in Fig.~\ref{couplingc}(a) and Fig.~\ref{couplingc}(b), respectively.  
We also found that $c^{\bar{\textrm{hy.}}}_{\bar{\omega}}$ with $f^{T}_{\bar{\omega}}=0$ shows stable Borel behavior 
and the plateau scale becomes moderate when $s_0^{} \geq 1.44~\textrm{GeV}^2$, which means that the $\bar{\omega}$ state 
can be interpreted as a hybrid state. 
Analogously, since $f^{T}_{{\bar{b}}_1 }$ with $ c^{\bar{\textrm{hy.}}}_{{\bar{b}}_1 }=0$ provides a plateau structure
and the plateau scale becomes moderate when $s_0^{} \geq3.24~\textrm{GeV}^2$, the one particle $b_1(1235)$ state would be
good enough to approximate the ${\bar{b}}_1$ state.

\section{Discussion and Conclusion}
\label{sec:summary}

%%%%%% FIG 7
\begin{figure}
\centering
\includegraphics[height=0.2\columnwidth]{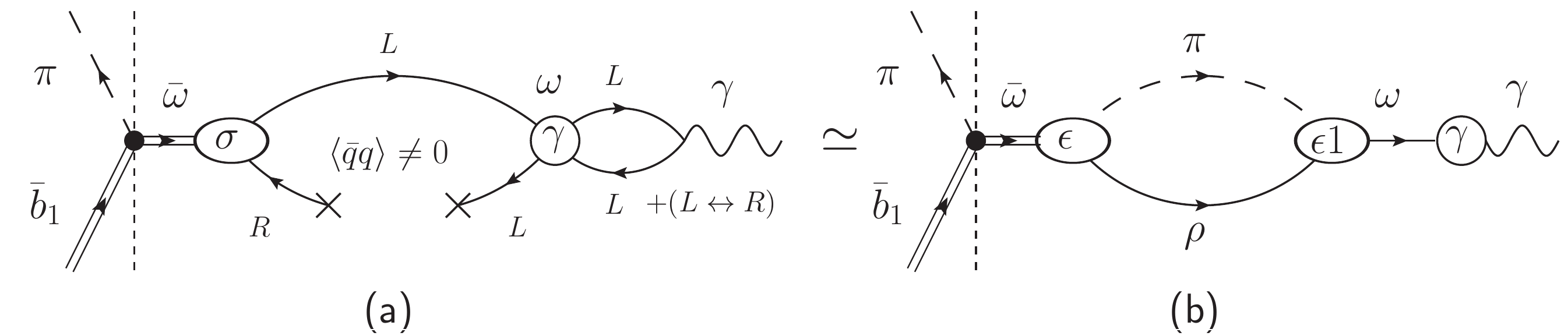}
\caption{The decay of $\bar{b}_1\rightarrow\pi \gamma$ through the $\omega(782)$ state, i.e., the VMD channel. 
(a) OPE description via QCD Lagrangian as inferred from Eq.~\eqref{b1gamma} and (b) hadron interaction
picture via anomalous coupling. 
In (a), the blobs with `$\sigma$' and `$\gamma$' represent the tensor and vector current representation, respectively, and
the vertex `$\epsilon$' and `$\epsilon1$' in (b) represent the anomalous hybrid state interpolation~\eqref{phencurrent} 
and the physical interaction vertex~\eqref{omegaac}, respectively.}
\label{tvcor}
\end{figure}

The $b_1(1235)$ meson state mostly decays into the $\pi\omega$ state by the strong interactions 
and the $\pi\gamma$ state by the electromagnetic interactions. 
As the VMD scenario is successful in explaining the decays and form factors of ground state hadrons, 
low-energy phenomenology usually assumes that the photon always sees hadrons through vector meson 
states~\cite{GS68, SS72a,Rosner81b, IYO89}. 
However, for the electromagnetic decays of the $b_1(1235)$ meson, the VMD hypothesis is not enough to explain 
the measured $\Gamma(b_1 \rightarrow \pi \gamma)$~\cite{Rosner81b, IYO89}.
In the present work, we explore the spin-1 mesonic states using the tensor representation for the purpose of searching for
a VMD-evading process for the $b_1$ radiative decay. 
Because of the mixing of helicity states, the corresponding currents are not generally conserved in their parity 
or isospin eigenspace. 
If $\mbox{SU(2)}_{L,R}$ flavor symmetry is assumed, the four different spin-1 mesonic states reside in the grand 
$\mbox{U(2)}_L \otimes \mbox{U(2)}_R$ symmetry group~\cite{CJ96}.

As chiral symmetry is spontaneously broken, the $\bar{b}_1$ state in the tensor representation decays into the chiral partner 
$\bar{\omega}$ state by emitting a pion as the Goldstone boson. 
From the QCD sum rules analysis, we find that the $\bar{b}_1$ state can be identified as the physical $b_1(1235)$ state, 
while it is problematic to identify the $\bar{\omega}$ state as the physical $\omega(782)$ meson. 
The $\bar\omega$ state has the same spin and flavor quantum numbers as the $\omega(782)$ but it is found to have a much higher mass. 
Comparing the weighted OPE invariant with the anomalous loop-like spectral structure, the $\bar{\omega}$ state can be identified 
as the hybrid state of the $\pi$ and $\rho$ mesons. 
As there is no observed $\omega$ resonance state in the mass of around $1~\textrm{GeV}$, the $\bar{\omega}$ state would be interpreted
as an intermediate virtual state.  
It is well known that the vector current $J^{0}_{\alpha }=\bar{q} T^0 \gamma_{\alpha} q$ couples strongly  to the vector meson state 
and the the resulting spectral sum rules successfully reproduce the $\omega(782)$ mass~\cite{SVZ79a, SVZ79b,SVZ79c, RRY85}. 
Since the isoscalar component of $J^{\textrm{e.m.}}_{\alpha }$ is same as $(e_0/3)J^{0}_{\alpha}$, the nonzero invariant of Eq.~\eqref{b1gamma3} 
can be understood as the overlap between the $\bar{\omega}$ hybrid state and the physical $\omega(782)$ state, which then becomes
the photon following the usual VMD scenario as depicted in Fig.~\ref{tvcor}~\cite{GS68, SS72a}.

\begin{figure}
\centering
\includegraphics[height=0.2\columnwidth]{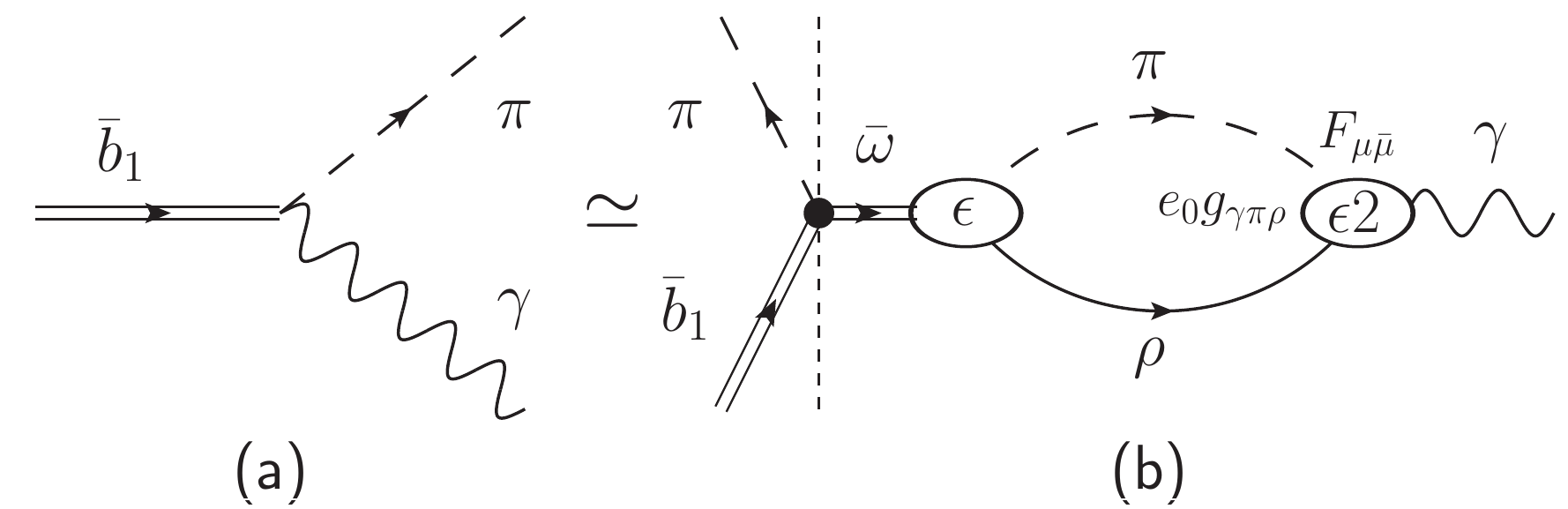}
\caption{(a) The $\bar{b}_1\rightarrow\pi \gamma$ decay not through the $\omega(782)$  state
(direct channel) and (b) phenomenological description for the direct coupling between $\bar{\omega}$ and $\gamma$. 
The vertex `$\epsilon$' and `$\epsilon2$' in (b) represent the anomalous hybrid state interpolation~\eqref{phencurrent} 
and the physical interaction vertex given in Eq.~\eqref{phvtx}, respectively.}
\label{direct}
\end{figure}

However, the loop structure of the $\bar{\omega}$ state suggests other mechanisms of radiative decays of the $b_1$.  
The  $\pi$-$\rho$ hybrid state can have an anomalous electromagnetic interaction as
\begin{align}
\mathcal{L}^{\epsilon2}_{\gamma \pi\rho}&=\frac{e_0g_{\gamma
\pi\rho}}{2m_{\rho}}\epsilon^{\mu \bar{\mu}\alpha \bar{\alpha}}F_{\mu\bar{\mu}}
\partial_{\alpha}\pi^a \rho_{\bar{\alpha}}^{a}\simeq\frac{e_0g_{\gamma
\pi\rho}}{m_{\rho}} F_{\mu\bar{\mu}}
\bar{\omega}^{\mu\bar{\mu}}. \label{phvtx}
\end{align}
If the $\pi$ and $\rho$ are on-shell, the vertex can be substituted with Eq.~\eqref{omegaac} by the VMD hypothesis~\cite{Meissner88}.
However, as most part of the loop phase space is off-shell, the direct photon coupling channel~$\eqref{phvtx}$ can be distinguished
from the VMD channel which could be the additional mechanism for the $b_1$ radiative decay as depicted in Fig.~\ref{direct}.

If the $b_1$ strongly couples to the tensor current, its decay is mediated mostly by the $\bar{\omega}$ state that is interpreted as
the $\rho$-$\pi$ hybrid state.
Then the major decay process of the $b_1$ meson would take the process of $b_1 \to \pi \bar{\omega} \to \pi \omega$, which
involves the anomalous $\omega\rho\pi$ interaction.
Because of the intermediate $\bar\omega$ state, the strong decay of the $b_1$ meson into two-pions, i.e., $b_1 \to \pi\pi\rho$, would
be enhanced enough to be measured.
However, because of the lack of experimental data, we cannot make a conclusion on the tensor representation of axial-vector mesons.
For example, the direct decay of $b_1 \to \pi\pi\pi\pi$ was reported in 1960s~\cite{ALMXY63}, which concluded only that 
$\Gamma(b_1 \to \pi^+\pi^+\pi^-\pi^0, \mbox{ direct})/\Gamma(b_1\to \omega\pi) < 0.5$.
More detailed information on the direct four-pion decays of the $b_1$ such as the precise value of the branching ratio and 
invariant mass distributions would be useful to understand the mechanisms of $b_1$ decay.
Therefore, precise measurement of the details of the $b_1$ decay will shed light on our understanding on the structure and properties of
axial-vector mesons and it can be performed at current experimental facilities.

%%%%%% FIG 8

\appendix

\section{\boldmath Tensor currents in $\mbox{SU(2)}_L \times \mbox{SU(2)}_R$ symmetry}
\label{appenixb}

\subsection{Transformation of tensor currents in the non-relativistic limit}

\begin{figure}
\centering
\includegraphics[width=0.46\columnwidth]{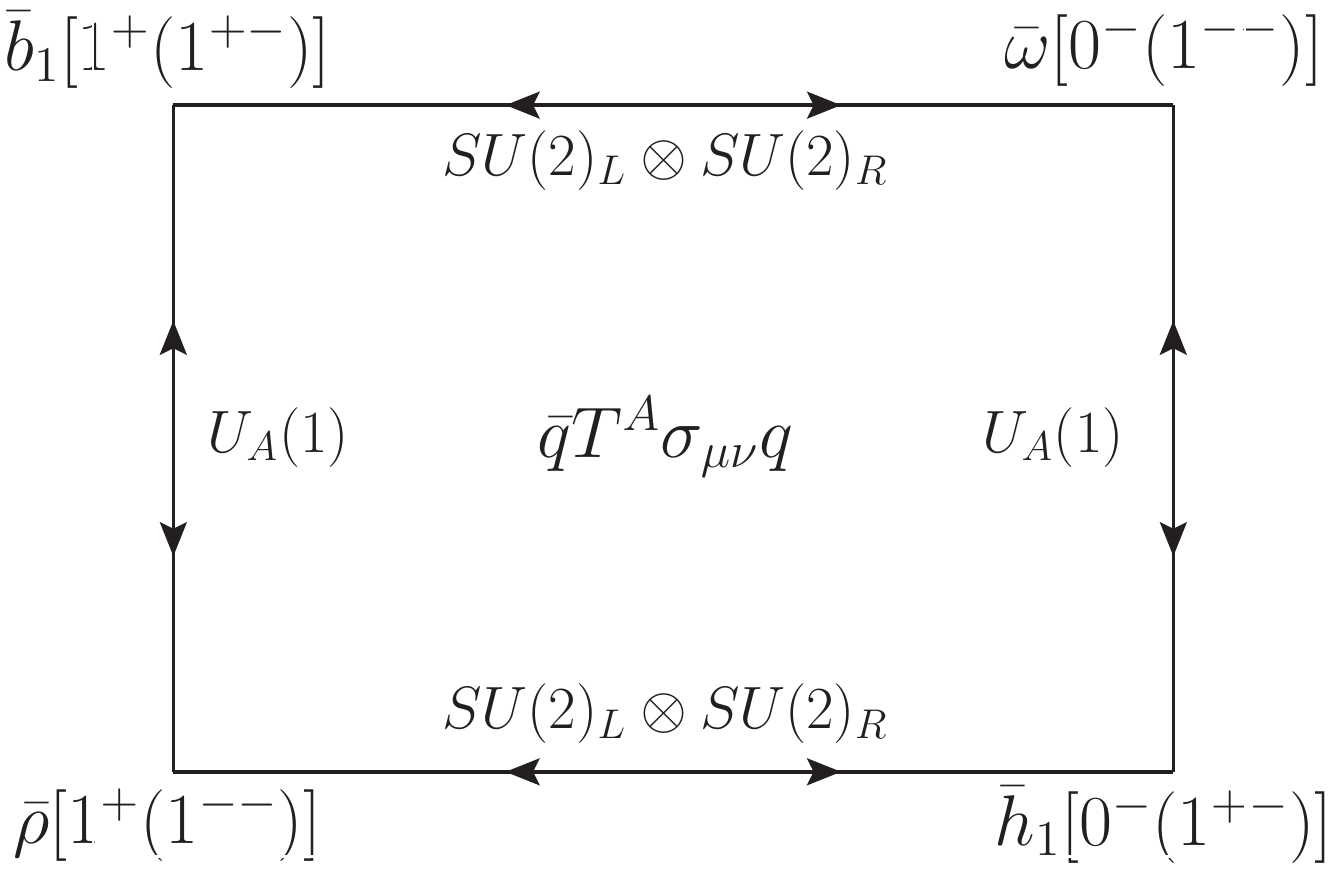}
\caption{Four different spin-1 mesonic states in tensor representation under  
$\mbox{U(1)} \times \mbox{SU(2)}_L \times \mbox{SU(2)}_R$. 
These constitute quartet elements of $\mbox{U(2)}_L \times \mbox{U(2)}_R$.}
\label{su2dia}
\end{figure}

The quantum numbers of the mesonic state are defined at  rest frame.
Equations~\eqref{n1}-\eqref{n4} show that each spin-1 state can be interpolated via tensor currents. 
Since the tensor current couples to both parity eigenstates, one can use dual tensor definition 
to describe a parity eigenstate in the opposite parity projection. 
For convenience, all the current is defined to
interpolate spin-1 mesonic states in parity-odd projection. 
The infinitesimal $\mbox{U}_A(1)$ transformation on the $\bar{b}_1$ current reads
\begin{eqnarray}
J^{\bar{b}_1,a}_{k} &=& \frac{1}{2}\epsilon_{ijk}^{} \bar{q}  T^a \sigma_{ij}^{} q  
\nonumber\\   
&\to& \frac{1}{2} \epsilon_{ijk}^{} \bar{q}  \left( I - i\alpha T^0 \gamma_5^{} \right) T^a
\sigma_{ij}^{} \left(I - i\alpha T^0 \gamma_5^{} \right) q  + O(\alpha^2)
\nonumber\\
&=& 
\frac{1}{2} \epsilon_{ijk}^{} \bar{q} T^a \sigma_{ij}^{} q  -i \alpha \left(-i \bar{q}  T^a \sigma_{0k}^{} q  \right) + O(\alpha^2) 
\nonumber \\ 
&=& J^{b_1,a}_{k} -i\alpha(-i J^{\rho,a}_{k} )+ O(\alpha^2),
\end{eqnarray}
where the corresponding commutation relations are summarized as
\begin{align}
[Q^0_5, J^{\bar{b}_1,a}_{k}(x)  ]& = -i J^{\bar{\rho},a}_{k}(x), \qquad 
[Q^0_5,J^{\bar{\rho},a}_{k}(x) ] = i J^{\bar{b}_1,a}_{k}(x).
\end{align}
Similarly, one can rotate quark fields via infinitesimal  $\mbox{SU}(2)_L \times \mbox{SU}(2)_R$ rotation as
\begin{align}
\left( 1 - i \theta^b T^b \right) q_L^{} \leftrightarrow q^\dagger_L \left(1+i\theta^b T^b \right),\\
\left( 1 - i\delta^b T^b \right) q_R^{} \leftrightarrow q^\dagger_R \left(1+i\delta^b T^b \right).
\end{align}
The transformation up to the leading order of $\theta^a$ and $\delta^a$ can be expressed 
in terms of explicit helicity states as
\begin{eqnarray}
J^{\bar{b}_1,a}_{k} &=& \frac{1}{2}\epsilon_{ijk}^{} \bar{q}  T^a \sigma_{ij}^{} q  
= \bar{q}_L^{} T^a \sigma_k^{} q_R^{} + \bar{q}_R^{} T^a \sigma_k^{} q_L^{}
\nonumber\\ &\to& 
\bar{q}_L ^{}T^a \sigma_k^{} q_R^{} + \bar{q}_R^{} T^a \sigma_k^{} q_L^{}
\nonumber\\ && \mbox{} 
+ i \theta^b \Bigg[ \bar{q}_L^{} \left(\frac{1}{4}\delta^{ab}\right) \sigma_k^{} q_R^{} 
+ \bar{q}_L^{} \left( \frac{i}{2}\epsilon^{bac}T^c \right) \sigma_k^{} q_R^{} \nonumber\\ &&\qquad\qquad
- \bar{q}_R^{} \left( \frac{1}{4}\delta^{ab} \right) \sigma_k^{} q_L^{} 
- \bar{q}_R^{} \left( \frac{i}{2}\epsilon^{abc}T^c \right) \sigma_k^{} q_L^{} \Bigg]
\nonumber\\ && \mbox{}
- i \delta^b \Bigg[ \bar{q}_L \left(\frac{1}{4}\delta^{ab}\right) \sigma_k^{} q_R^{} 
+ \bar{q}_L \left(\frac{i}{2} \epsilon^{abc}T^c \right) \sigma_k^{} q_R^{}  \nonumber\\ &&\qquad\qquad
-\bar{q}_R \left(\frac{1}{4}\delta^{ab}\right) \sigma_k^{} q_L^{} 
- \bar{q}_R \left(\frac{i}{2}\epsilon^{bac}T^c \right) \sigma_k^{} q_L^{} \Bigg]  \mbox{}
+ O(\theta^2,\delta^2)
\nonumber\\
&\simeq& \frac{1}{2}\epsilon_{ijk}^{} \bar{q}  T^a \sigma_{ij}^{} q 
+ \frac{i (\theta^b+\delta^b )}{2} \left( -i\epsilon^{abc} \frac{1}{2} \epsilon_{ijk}^{} \bar{q}  T^c\sigma_{ij}^{} q   \right)
+ \frac{i (\theta^a-\delta^a )}{2} \left( -i\bar{q}  T^0 \sigma_{0k}^{} q  \right)
\nonumber\\
&=& J^{\bar{b}_1,a}_{k} + \frac{i (\theta^b+\delta^b )}{2} \left(-i \epsilon^{abc} J^{\bar{b}_1,c}_{k}  \right)  
+ \frac{i  (\theta^a-\delta^a )}{2} \left(-i J^{\bar{\omega}}_{k}  \right),
\end{eqnarray}
where the rotation directions `$\theta^a=\delta^a$' and `$\theta^a=-\delta^a$' correspond to 
$Q^a_v=\int d^3x\, q^\dagger(x) T^a q(x)$ and $Q^a_5=\int d^3x\, q^\dagger(x) T^a \gamma_5^{} q(x)$,
respectively.  
The corresponding commutation relations read
\begin{align}
[Q^a_5, J^{\bar{b}_1,b}_{k}(x) ] = -i \delta^{ab}J^{\bar{\omega}}_{k}(x),
\qquad
[Q^a_5, J^{\bar{\omega}}_{k}(x) ]  =  i J^{\bar{b}_1,a}_{k}(x),
\qquad 
[Q^a_v, J^{\bar{b}_1,b}_{k}(x) ] = -i\epsilon^{abc}J^{\bar{b}_1,c}_{k}(x).
\end{align}
Similar
commutation relations can be obtained for other currents as
\begin{align}
[Q^0_5, J^{\bar{\omega}}_{k}(x) ] = -i J^{\bar{h}_1}_{k}(x),
\qquad
[Q^a_5, J^{\bar{\rho},b}_{k}(x) ] = i \delta^{ab}J^{\bar{h}_1}_{k}(x),
\qquad
[Q^a_v, J^{\bar{\rho},b}_{k}(x) ] = -i \epsilon^{abc}J^{\bar{\rho},c}_{k}(x).
\end{align}
The mixing properties are diagrammatically described in Fig.~\ref{su2dia}.

%%%%%% FIG 9

\subsection{Transformation of tensor currents in relativistic generalization}

The covariant currents are defined as $J^0_{\mu\bar{\mu}} \equiv \bar{q} T^0 \sigma_{\mu\bar{\mu}}^{} q$
and $\tilde{J}^a_{\mu\bar{\mu}} \equiv -\frac{1}{2} \epsilon_{\mu\bar{\mu}\alpha\beta}^{} \bar{q}  T^a
\sigma^{\alpha\beta} q$ to describe the $\bar{\omega}$ and $\bar{b}_1$ states, respectively, in the parity-odd projection 
at boosted frame. 
The flavor axial charge commutation relations are
\begin{eqnarray}
\left [Q^a_5,J^{0}_{\mu\bar{\mu}}(x)\right] &\to& 
-\bar{q}(x) T^a \gamma_5^{} \sigma_{\mu\bar{\mu}}^{} q(x)
\nonumber \\ 
&=&
-\frac{i}{2} \epsilon_{\mu\bar{\mu}\alpha\beta}^{} \bar{q}(x) T^a \sigma^{\alpha\beta} q(x) 
= i \tilde{J}^{a}_{\mu\bar{\mu}}(x),
\\
 \left[Q_5^a, \tilde{J}^{b}_{\mu\bar{\mu}}(x) \right] &\to&
\frac{1}{2} \epsilon_{\mu\bar{\mu}\alpha\beta}^{} \delta^{ab} \bar{q}(x) T^0 \gamma_5^{} \sigma^{\alpha\beta} q(x) 
\nonumber\\
&=&
\frac{1}{2} \epsilon_{\mu\bar{\mu}\alpha\beta}^{} \frac{i}{2} \epsilon^{\alpha\beta\nu\bar{\nu}}
\delta^{ab} \bar{q}(x) T^0  \sigma_{\nu\bar{\nu}}^{} q(x)
=-i\delta^{ab}J^{0}_{\mu\bar{\mu}}(x).
\end{eqnarray}
Similarly, the other commutation relations can be obtained in covariant generalization as
\begin{align}
& [Q^a_5,J^{b}_{\mu\bar{\mu}}(x)] = -i\delta^{ab}\tilde{J}^{0}_{\mu\bar{\mu}}(x) ,
\quad &
[Q^a_5,\tilde{J}^{0}_{\mu\bar{\mu}}(x)] = iJ^{a}_{\mu\bar{\mu}}(x) ,\\
& [Q^0_5,\tilde{J}^{a}_{\mu\bar{\mu}}(x)] =  -i J^{a}_{\mu\bar{\mu}}(x), 
& [Q^0_5,J^{a}_{\mu\bar{\mu}}(x)] = i \tilde{J}^{a}_{\mu\bar{\mu}}(x),\\
& [Q^0_5,J^{0}_{\mu\bar{\mu}}(x)]=-i \tilde{J}^{0}_{\mu\bar{\mu}}(x),
& [Q^0_5,\tilde{J}^{0}_{\mu\bar{\mu}}(x)] = i J^{0}_{\mu\bar{\mu}}(x).
\end{align}
These lead to the same diagrammatical summarization presented in Fig.~\ref{su2dia}.

\section{Technical remarks on sum rules analysis}
\label{appenixc}

\subsection{Four-quark condensates}

%%%%%% FIG 10
\begin{figure}
\centering
\includegraphics[height=0.2\columnwidth]{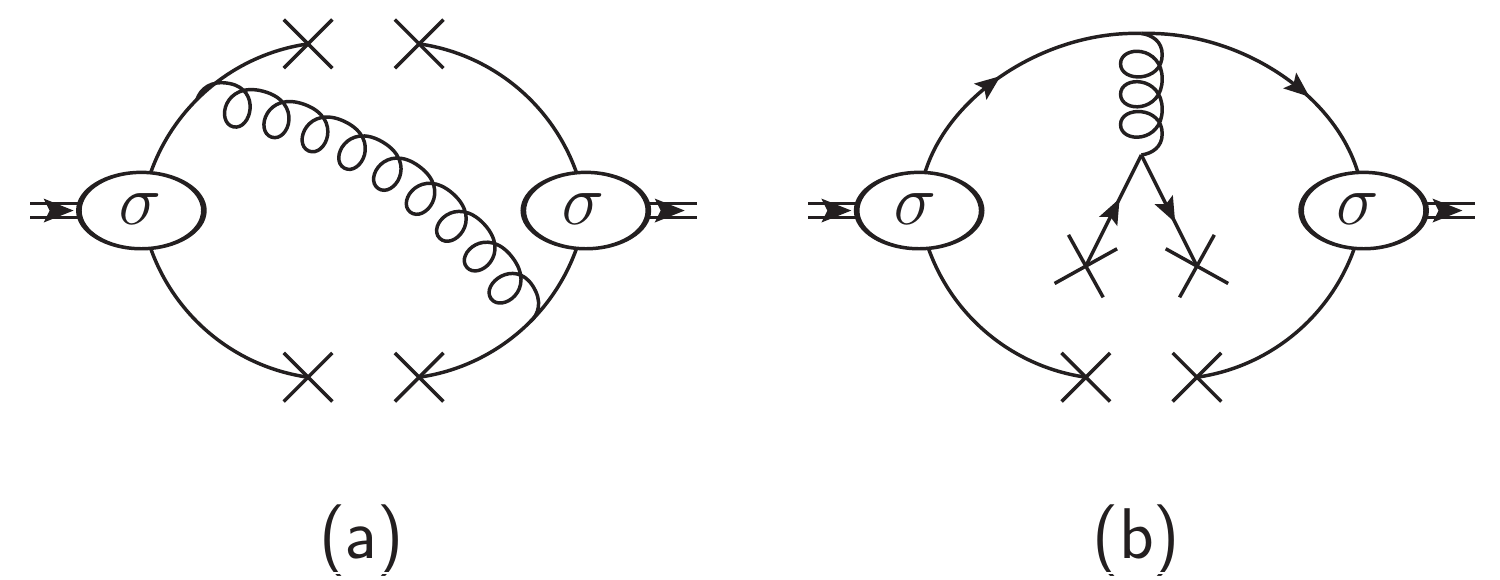}
\caption{Diagrammatical description of the four-quark contributions
in the OPE of correlator~\eqref{tcorr1}. (a) contribution to the
invariant~\eqref{4qdisc} and (b) the
invariant~\eqref{4qconn}. 
The blob with `$\sigma$' represents the tensor current representation.}
\label{41disc}
\end{figure}

\begin{figure}
\centering
\includegraphics[height=0.2\columnwidth]{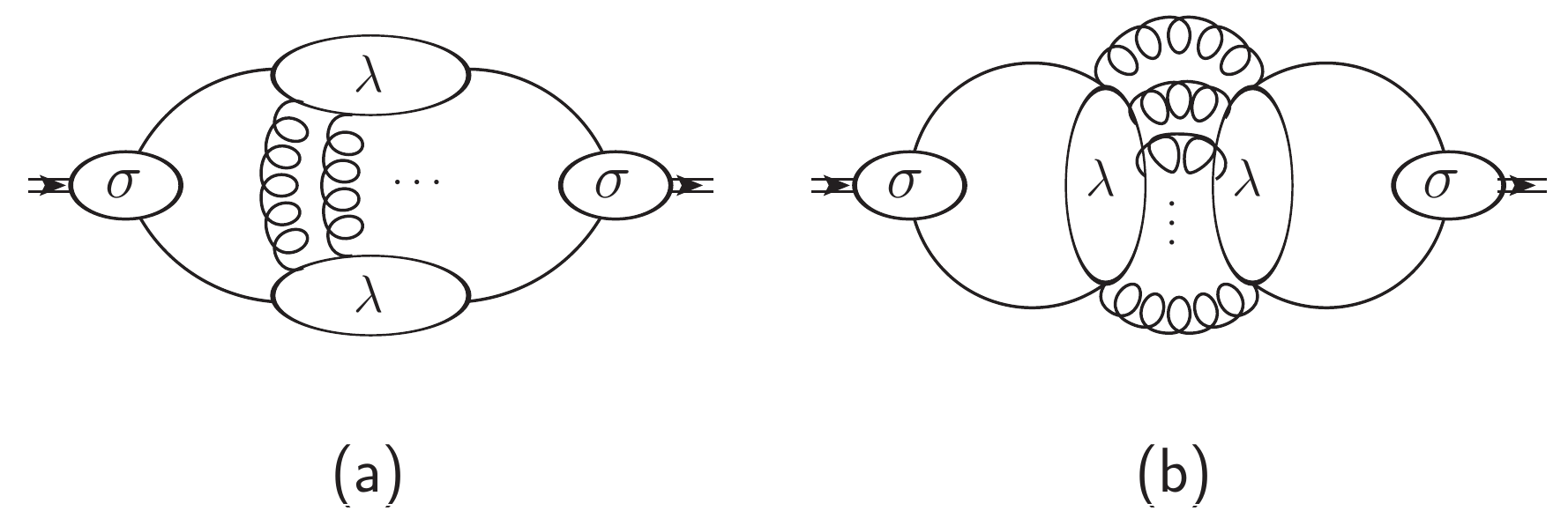}
\caption{Correlation types of the mesonic quantum numbers:  
(a) connected and (b) disconnected correlation of Dirac eigenmodes. 
Here, `$\sigma$' in the blob and `$\lambda$' in the quark correlation represent the tensor current representation and the Dirac eigenmode, respectively. }  
\label{41disc2}
\end{figure}

In the OPE of the tensor current correlator, the four-quark condensate terms are the
lowest mass-dimensional quark contribution.
These condensates determine the asymptotic behavior of the Borel curves and distinguish parity eigenstates. 
The condensates in Eqs.~\eqref{factcond1} and \eqref{factcond2} give huge contribution to the
sum rules because the Wilson coefficient of the invariant term of Eq.~\eqref{4qdisc} is dominant over the others. 
The corresponding diagrammatical description can be found in Fig.~\ref{41disc}(a).  
Unfortunately, the values of these four-quark condensates are unknown at present, and are thus estimated,
for example, by adopting the vacuum saturation hypothesis. 
However, the usual factorization scheme does not reflect all the possible contributions in the condensates. 
In the condensates defined in Eqs.~\eqref{factcond1} and \eqref{factcond2}, isospin and color matrices appear 
in the condensates, and the usual factorization scheme only allows `quark connected' correlations as in Fig.~\ref{41disc2}(a) 
when there are isospin matrices in the quark bilinears.
Although isovector terms cannot have `quark disconnected' contribution [Fig.~\ref{41disc2}(b)], the isoscalar terms 
can have contributions even if they are colored. 
One can consider the non-local generalization of meson interpolating current,
\begin{align}
M^A_{\mu\bar{\mu}}(y,x;c) & =   \bar{q} (y ) T^A \sigma_{\mu\bar{\mu}}^{}
\exp \left[ig \int_{x}^{y} dz^\nu A_\nu(z) \right] q(x).
\label{nlo}
\end{align}
Then, in the very small but nonzero space-time separation $\epsilon = \vert y-x \vert \neq 0 \ll1$ which can be 
allowed in the relevant hadron size for the quark correlation average, gauge corrections are needed to fix the initial 
and final interpolated states on the equivalent gauge orbit [Fig.~\ref{41disc2}(b)]. 
Therefore, if the pair of colored isoscalar bilinear appears in totally color blind configuration, there is no reason to
discard disconnected contribution. 
Considering that the relative sign between two contributions is negative, the magnitude of the isoscalar parameter is assigned 
to be smaller than that of the isovector parameter: $| \alpha_{0} | < | \alpha_{a}|$, $| \beta_{0} | < | \beta_{a} |$.

%%%%%% FIG 11

We now consider the condensates containing pseudoscalar type bilinear as in Eq.~\eqref{factcond2}. 
Since $\psi_0$ and $\gamma_5^{} \psi_0$ reside in the same eigenspace $(\lambda=0)$, the vacuum average of
pseudoscalar bilinear does not vanish in the nontrivial gauge configuration~\cite{Coleman}:
\begin{equation}
\braket{ \bar{q} \gamma_5^{} q }_{{\nu}}  =  - \lim_{m_q \rightarrow 0} 
\Braket{ \sum_{\lambda} \frac{ \psi_{\lambda}^\dagger \gamma_5^{} \psi_{\lambda}  }{m_q-i\lambda}}
= - \pi(n_{R}^{} - n_{L}^{}) \equiv \pi \nu ,
\label{pseudo}
\end{equation}
where $\nu$ denotes the winding number, and the inner product between the modes in different eigenspace vanishes.  
If all the gauge configuration is summed up, the vacuum average can be written as
\begin{equation}
\braket{ \bar{q} \gamma_5^{} q}   =  \sum_{\nu} e^{i \nu \theta} \braket{\bar{q} \gamma_5^{} q}_\nu = 
2\pi i \sum_{\nu=0}^{\infty} \nu \sin{\nu \theta}  \simeq 0,
\label{pseudosum}
\end{equation}
where $\theta$ is the QCD vacuum angle. 
However,  for the quark correlation in the four-quark vacuum average, the nontrivial topological contribution 
should appear in terms of $\nu^2$ and would not vanish in the summation. 
In this study, the nontrivial contribution is assumed to appear as an alternating series of
$\nu^2$ in Eq.~\eqref{factcond2} while it does not in Eq.~\eqref{factcond1}, which leads to
$\vert \alpha_A \vert > \vert \beta_A \vert$.

The condensate~\eqref{factcond3} contributes to the invariant piece~\eqref{4qconn} whose diagrammatical description 
is given in Fig.~\ref{41disc}(b). 
As the vector bilinear is obtained from the field equation of the gluons, at least the sign of the usual factorization
estimation has been justified in Euclidean space~\cite{SVZ79a}. 
Therefore we take the usual vacuum saturation value of $\gamma=1$.  
Furthermore,  its contribution in Eq.~\eqref{4qconn} is  much smaller than those from the other four-quark condensates 
so that a slightly different value for gamma will not change the main result of the present work.

\subsection{Supplementary remarks for the sum rules with $\alpha_A < 0$, $\beta_A<0$}

%%%%%% FIG 12
\begin{figure}
\centering
\includegraphics[height=5.9cm]{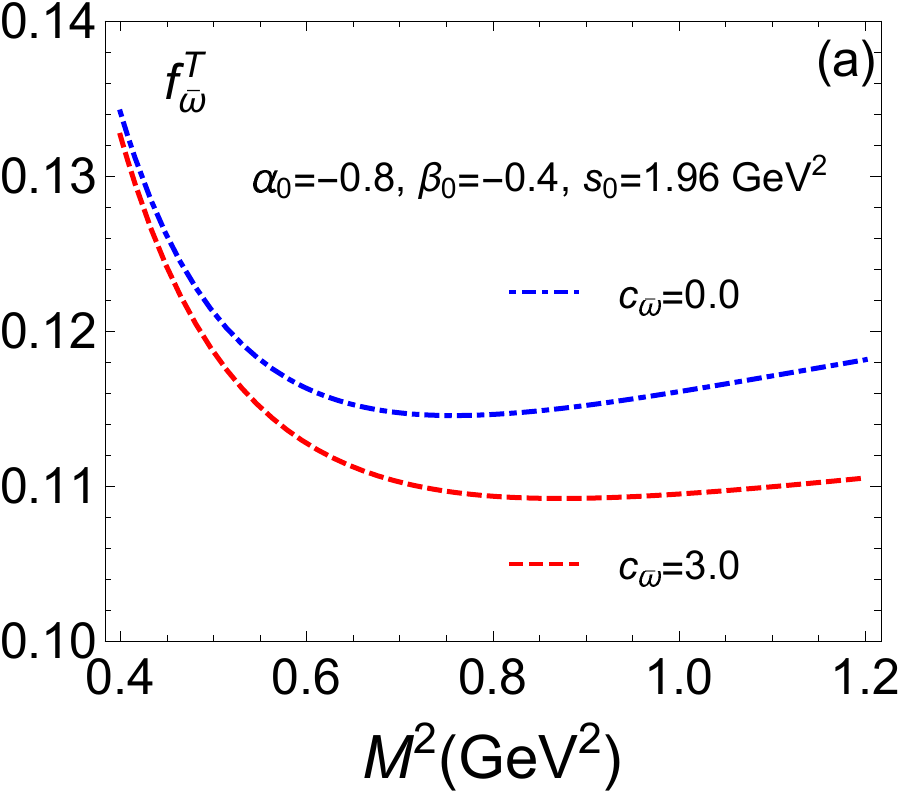} \qquad
\includegraphics[height=5.9cm]{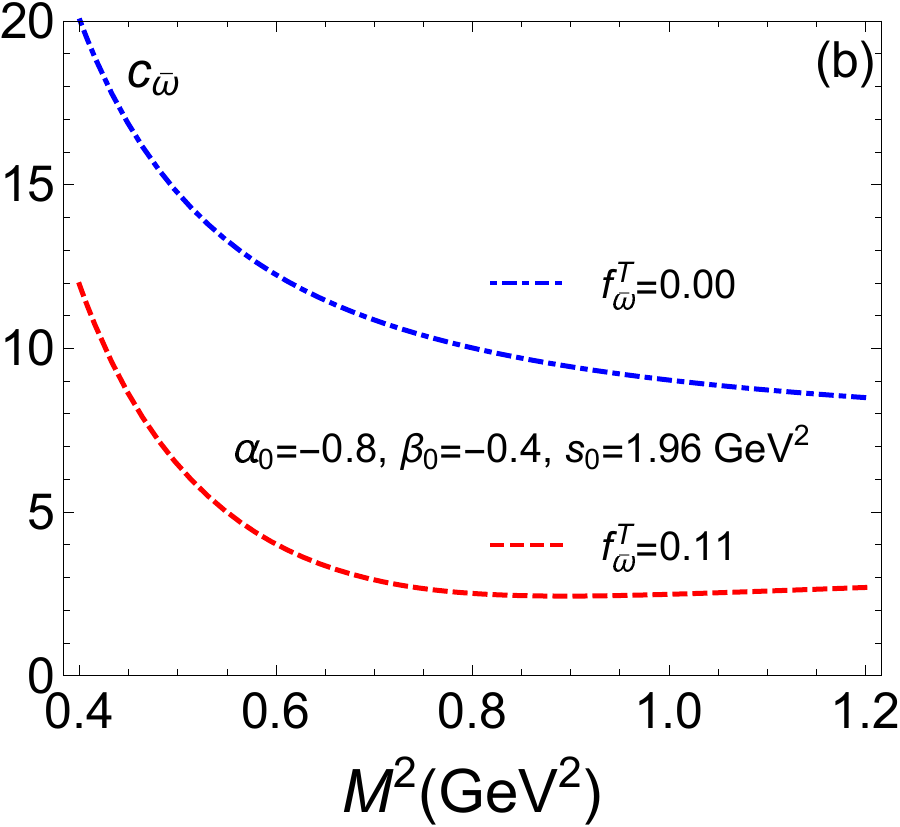}  
\caption{Borel curve (a) for the meson pole residue $f^{T}_{\bar{\omega} }$ and 
(b) for the tensor current-hybrid coupling $ c^{\bar{\textrm{hy.}}}_{\bar{\omega} }$ with reversed parameter set $(\alpha_0=-0.8,~\beta_0=-0.4)$.}\label{f14}
\end{figure}

\begin{figure}
\centering
\includegraphics[height=5.8cm]{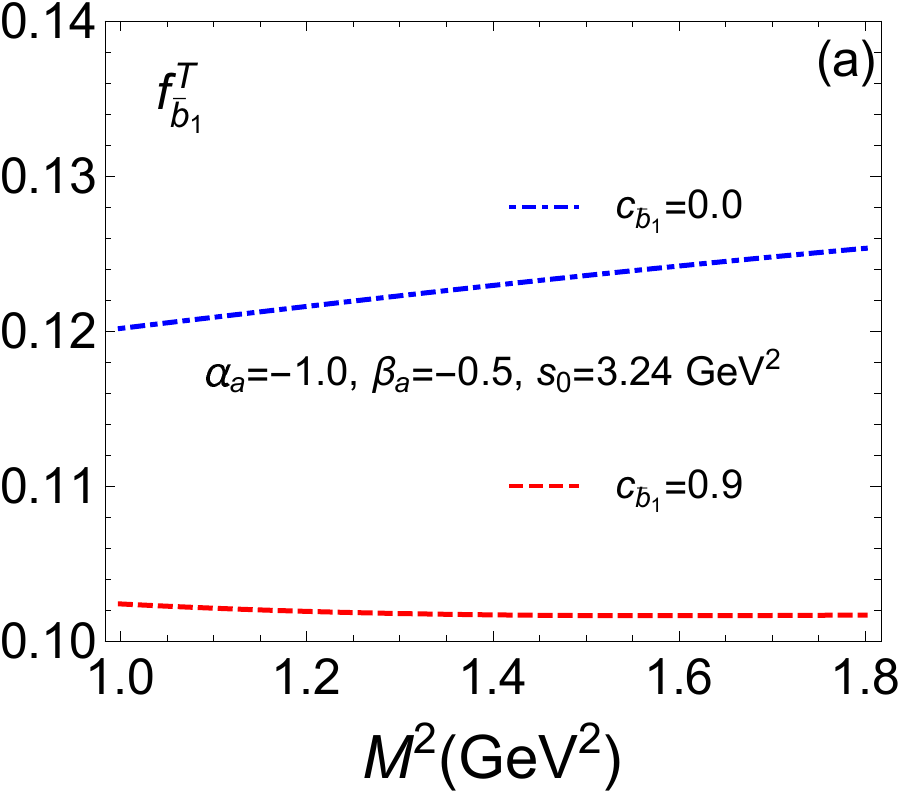} \qquad
\includegraphics[height=5.8cm]{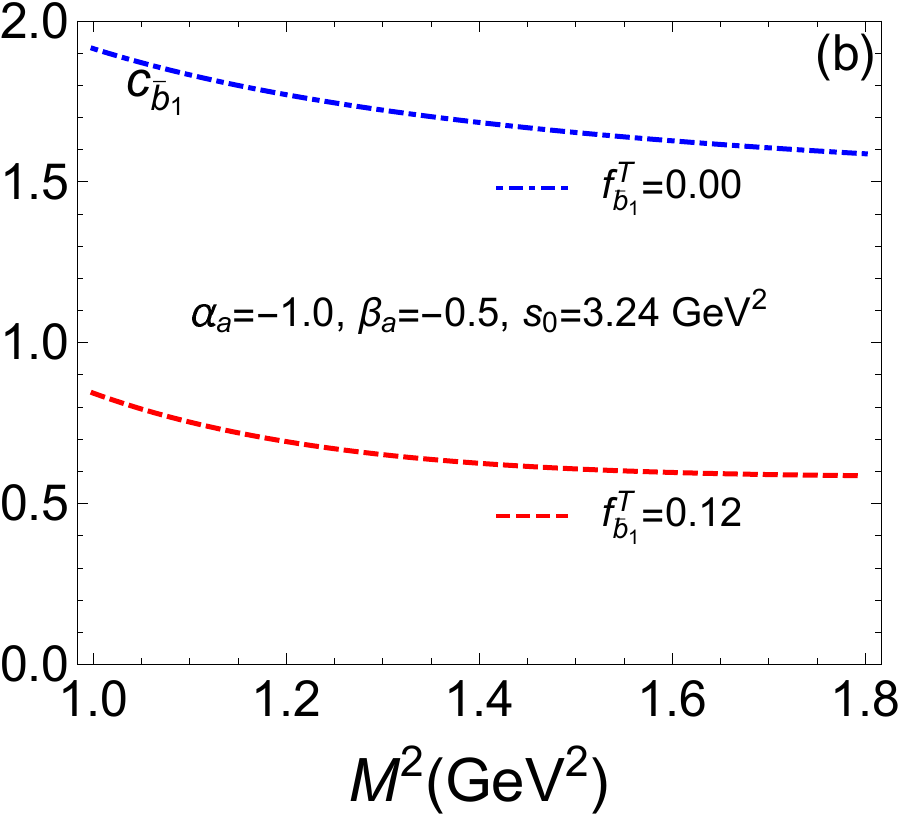}  
\caption{Borel curve (a) for the meson pole residue $f^{T}_{\bar{b}_1 }$ and 
(b) for the tensor current-hybrid coupling $ c^{\bar{\textrm{hy.}}}_{\bar{b}_1 }$ with reversed parameter set 
$(\alpha_a=-1.0,~\beta_a= -0.5)$. }
\label{f15}
\end{figure}

In the previous arguments, the proper sets $(\alpha_0=0.8$, $\beta_0=0.4)$ and $(\alpha_a=1.0$, $\beta_a=0.5)$ have
been used for Borel sum rules for the $\bar{\omega}$ and $\bar{b}_1$ states, respectively. 
If we use the reversed sign of the parameter set, the Borel curves for $f^{T}_{\mp}$ and $c^{\bar{\textrm{hy.}}}_{\mp}$ are obtained 
as shown in Fig.~\ref{f14} and Fig.~\ref{f15}, respectively. 
In this case, $f^{T}_{\bar{\omega} }$ shows plateau-like behavior even in the one-particle pole limit 
($c^{\bar{\textrm{hy.}}}_{\bar{\omega} }=0$). 
In the optimized condition, the stable Borel curves provide $f^{T}_{\bar{\omega} }=0.11$ and 
$c^{\bar{\textrm{hy.}}}_{\bar{\omega} }=2.0$ as plateau values. 
It is found that the magnitude of $f^{T}_{\bar{\omega} }$ reduces to $90\%$ when we vary $c^{\bar{\textrm{hy.}}}_{\bar{\omega} }$
from $0$ to $2.0$, while $c^{\bar{\textrm{hy.}}}_{\bar{\omega} }$ becomes $20\%$ of its original value when we change 
$f^{T}_{\bar{\omega} }$ from $0$ to $0.11$. 
This means that the $\bar{\omega}$ state becomes dominated by the one-particle mesonic state 
with the set of $(\alpha_0^{}=-0.8$, $\beta_0^{}=-0.4)$.

However, this tendency becomes unclear in the case of the $\bar{b}_1$ analysis. 
As one can find in Fig.~\ref{f15}, $f^{T}_{\bar{b}_1 }$ and $ c^{\bar{\textrm{hy.}}}_{\bar{b}_1 }$ do not show stable behavior 
with $ c^{\bar{\textrm{hy.}}}_{\bar{b}_1 }=0$ and $f^{T}_{\bar{b}_1 }=0.00$, respectively.  
In the optimized condition, even though the Borel curve for $f^{T}_{\bar{b}_1 }$ strongly depends on the continuum threshold, 
the scale of each coupling strength becomes about $80\%$ and $45\%$ of their original magnitudes when we vary 
$c^{\bar{\textrm{hy.}}}_{\bar{b}_1 }$ from $0$ to $0.9$ and $f^{T}_{\bar{b}_1 }$ from $0$ to $0.12$, respectively. 
This observation then leads us to the conclusion that the $\bar{b}_1$ state is dominated by the one-particle meson state 
within the set of  $(\alpha_a=-1.0$, $\beta_a= -0.5)$. 
This result leads to the usual VMD scenario ($b_1 \rightarrow \pi\omega  \rightarrow  \pi\gamma $).

Considering that the Borel curve for $f^{T}_{\bar{b}_1 }$ is highly dependent of the parameter set, 
it is possible that the non-negligible part of $\bar{b}_1$ could be a  $\pi$-$\omega$ hybrid state.
However, this kind of hybrid state leads to a stable $\omega(782)$ state after pion breaking, 
where the VMD scenario for the radiative decay is still valid.

%%%%%% FIG 13
\subsection{Borel sum rules}

By assuming regularity of the correlation function, the invariants can
be written in dispersion relation as
\begin{equation}
\Pi_i(q^2)=\frac{1}{2\pi i}\int^\infty_{0} d s \frac{\Delta\Pi_i(s)}{s-q^2} +P_n(q^2), 
\label{vaccord}
\end{equation}
where $P_n(q^2)$ is the finite order polynomial in $q^2$, which comes from the integration 
on the circle of contour on complex plane and the discontinuity
$\Delta\Pi_i(s)\equiv\lim_{\epsilon\rightarrow0^{+}}[\Pi_i(s+i\epsilon)-\Pi_i(s-i\epsilon)]=2i\,\textrm{Im}[\Pi_i(s+i\epsilon)]$
is defined on the positive real axis. 
Then all the possible physical states are contained in this discontinuity. 
In the phenomenological point of view, the invariant can be assumed to have a pole and continuum structure as
\begin{align}
\Delta\Pi_i(s)=\Delta\Pi_i^{\textrm{pole}}(s)+\theta(s-s_0^{})\Delta\Pi_i^{\textrm{OPE}}(s),\label{disca}
\end{align}
where $s_0^{}$ represents the continuum threshold. 
To suppress the continuum contribution, the weight function $W(s)=e^{-s/M^2}$ may be used as
\begin{align}
\mathcal{W}_M[\Pi_i(q^2)]=\frac{1}{2\pi i}\int^\infty_{0} d s~ e^{-s/M^2}
 \Delta\Pi_i(s).\label{borelv}
\end{align}
The corresponding differential operator $\mathcal{B}$ can be defined as
\begin{align}
\mathcal{B}[f(-q ^2) ]\equiv \lim_{\substack{-q ^2,n\rightarrow\infty\\-q ^2/n=M^2}} 
\frac{(-q ^2)^{n+1}}{n!} \left(\frac{\partial}{\partial q ^2}\right)^n f(-q^2).
\label{boreltv}
\end{align}
By making use of this operator in Eq.~\eqref{vaccord}, one obtains
\begin{align}
\mathcal{W}_M[\Pi_i(q^2)]&=\frac{1}{2\pi i}\int^\infty_{0} d s~ e^{-s/M^2}
 \Delta\Pi_i(s) =\mathcal{B}[\Pi_i(q^2)],\label{borelv1}
\end{align}
where following relation has been used:
\begin{align}
 \mathcal{B}\left[\frac{1}{s-q^2}\right] =e^{-s/M^2}.
\end{align}
The residues located after the continuum threshold with finite $s_0$
can be subtracted as
\begin{align}
\mathcal{W}_M^{\textrm{subt.}}[\Pi_i(q^2)]= \frac{1}{2\pi i}\int^{s_0^{}}_0 d s\, e^{-s/M^2}
 \Delta\Pi_i(s)=\mathcal{B}[\Pi_i(q^2)]_{\textrm{subt.}}.
\end{align}

Using the following integral identities,
\begin{align}
\int_{s_0^{}}^\infty ds\, e^{-s/M^2}&=M^2e^{-s_0^{} /M^2},\\
\int_{s_0^{}}^\infty ds\, se^{-s/M^2}&=(M^2)^2e^{-s_0^{} /M^2}\left(s_0^{} /M^2+1\right),\\
\int_{s_0^{}}^\infty ds\, s^2e^{-s/M^2}&=(M^2)^3e^{-s_0^{} /M^2}\left(s_0^{2}/2M^4+s_0^{} /M^2+1\right),
\end{align}
the subtraction of continuum contribution for the OPE side can be summarized as
\begin{align}
E_0(s_0)&\equiv1-e^{-s_0^{} /M^2},\\
E_1(s_0)&\equiv1-e^{-s_0^{} /M^2}\left(s_0^{} /M^2+1\right),\\
E_2(s_0)&\equiv1-e^{-s_0^{} /M^2}\left(s_0^{2}/2M^4+s_0^{} /M^2+1\right).
\end{align}
These results are used when $E_n$ is multiplied to all $(M^2)^{n+1}$ terms in $\mathcal{B}[\Pi_i(q^2)]$.
This weighting scheme and subsequent spectral sum rules are known as the Borel transformation and Borel sum rules, respectively.

\acknowledgments
\newblock
This work was supported by the 
National Research Foundation of Korea under Grant Nos.
NRF-2017R1D1A1B03033685 (K.S.J.), NRF-2016R1D1A1B03930089 (S.H.L.), 
and NRF-2015R1D1A1A01059603 (Y.O.).

\end{document}